\documentstyle[epsfig]{mn}
\newif\ifAMStwofonts
\def\aj{{AJ}}			
\def\araa{{ARA\&A}}		
\def\apj{{ApJ}}			
\def\apjl{{ApJ}}		
\def\apjs{{ApJS}}

\def\aap{{A\&A}}		
		
\def\aaps{{A\&AS}}

\def\mnras{{MNRAS}}

\def\pasa{{PASA}}	
\def\pasp{{PASP}}

\def\nat{{Nature}}

\def\fcp{{Fund.~Cosmic~Phys.}}

\title[Star formation properties of local galaxies]
{Star formation properties of UCM galaxies}
\author[A. Gil de Paz {\rm et al.}]
       {A. Gil de Paz$^1$, A. Arag\'on-Salamanca$^{2,3}$,
       J. Gallego$^1$, A. Alonso-Herrero$^4$, \newauthor
       J. Zamorano$^1$ and G. Kauffmann$^5$\\ $^1$Departamento de
       Astrof\'{\i}sica, Facultad de F\'{\i}sicas, Universidad
       Complutense, E-28040 Madrid, Spain\\ $^2$Institute of
       Astronomy, Madingley Road, Cambridge CB3 0HA,
       England\\$^3$School of Physics and Astronomy, University of
       Nottingham, Nottingham, NG7 2RD, England\\$^4$Steward
       Observatory, The University of Arizona, Tucson AZ 85721, USA\\
       $^5$Max-Planck-Institut f\"ur Astrophysik, D-85740 Garching bei
       M\"unchen, Germany\\ e-mail: gil@astrax.fis.ucm.es (AGdP) }
\date{Accepted ---.
      Received ---;
      in original form ---}

\pagerange{\pageref{firstpage}--\pageref{lastpage}}
\pubyear{1999}

\begin{document}
\hyphenation{IIIaF}

\maketitle

\label{firstpage}

\begin{abstract}
\label{abstract} 

We present new near-infrared $J$ and $K$ imaging data for 67 galaxies
from the Universidad Complutense de Madrid survey used to
determine the SFR density of the local universe by Gallego et al.
(1995).  This is a sample of local star-forming galaxies with redshift
lower than 0.045, and they constitute a representative subsample of the
galaxies in the complete UCM survey. From the new data, complemented with our
own Gunn-$r$ images and long-slit optical spectroscopy, we have
measured integrated $K$-band luminosities, $r-J$ and $J-K$ colours, and
H$\alpha$ luminosities and equivalent widths.  Using a maximum
likelihood estimator and a complete set of evolutionary synthesis
models, these observations have allowed us to estimate the strength of
the current (or most recent) burst of star formation, its age, the
star-formation rate and the total stellar mass of the galaxies. An
average galaxy in the sample has a stellar mass of
5$\times$10$^{10}$\,M$_{\sun}$ and is undergoing (or recently
completed) a burst of star formation involving about 2~per cent of its
total stellar mass. We have identified two separate classes of
star-forming galaxies in the UCM sample:  low luminosity, high
excitation galaxies (HII-{\it like}) and relatively luminous spirals
galaxies (starburst disk-{\it like}).  The former show higher {\it
specific} star formation rates (SFR per unit mass) and burst strengths,
and lower stellar masses than the latter. With regard to their {\it
specific} star formation rates, the UCM galaxies are intermediate
objects between normal quiescent spirals and the most extreme HII
galaxies.
\end{abstract}


\section{Introduction}
\label{introduction}

The study of the evolution of the Star Formation Rate (SFR) of
individual galaxies and the SFR history of the Universe has
experienced considerable progress recently (see, e.g., Madau,
Dickinson \& Pozzeti 1998 and references therein). These are key
observables needed to extend our understanding of galaxy formation and
evolution. In the last few years, the combination of very deep
ground-based and HST multi-band imaging with deep spectroscopic
surveys carried out with 4-m and 10-m class telescopes has allowed the
sketching of the SFR history of the Universe up to $z>4$ (see, e.g.,
Madau et al. 1998 and references therein).

A great deal of effort has been devoted to both observational and
theoretical studies of star-forming objects and their evolution with
look-back-time. Deep imaging and spectroscopy of faint galaxies at
intermediate and high redhsifts have yielded vast amounts of
quantitative information in this field (Lilly et al. 1995; 1998 and
references therein; Driver, Windhorst \& Griffiths 1995; Steidel et
al. 1996; Lowenthal et al.  1997; Hammer et al. 1997; Hu, Cowie \&
McMahon 1998; see Ellis 1997 for a recent comprehensive
review). Although substantial uncertainties still exist, a reasonably
coherent picture is emerging. The Star Formation Rate density of the
universe was probably about an order of magnitude higher in the past
than it is now, perhaps peaking at $z\sim1$--$2$ (e.g., Gallego et
al. 1995; Madau et al. 1996; Connolly et al. 1997; Madau et al. 1998).
These observational results seem to be in good agreement with the
predictions of recent theoretical models of galaxy formation (Pei \&
Fall 1995; Baugh et al. 1998; Somerville, Primack \& Faber 1999), although
the question of whether the SFR density decreased beyond $z\sim2$ is
still a matter of intense debate (Hu et al. 1998; Somerville et al. 
1999; Hughes et al. 1998; Barger et al. 1998; Steidel et
al. 1999).

Given the large redshift range covered by these studies, different SFR
indicators have perforce been used, all of which have different
calibrations, selection effects and systematic uncertainties.  These
indicators include emission line luminosities (e.g., H$\alpha$,
H$\beta$, [OII]$\lambda3727$\AA), blue and ultraviolet fluxes,
far-infrared and sub-mm fluxes, etc (see, e.g., Gallego et al. 1995;
Rowan-Robinson et al. 1997; Tresse \& Maddox 1998; Glazebrook et al.
1999; Treyer et al. 1998; Madau et al. 1996; Connolly et al. 1997;
Hughes et al. 1998; Barger et al. 1998; see also Charlot 1998 and
Kennicutt 1992).  It would be highly desirable to use the same SFR
indicator at all redshifts, so that the problems related to different
selection effects and systematics could be avoided. It is widely
accepted that the H$\alpha$ is one of the most reliable measurements
of the current star formation rate ({\it modulo\/} the IMF; see, e.g.,
Kennicutt 1992).  Several groups have used the H$\alpha$ line to
estimate SFRs at different redshifts, from the local universe to
beyond $z=1$ (Gallego et al. 1995; Tresse \& Maddox 1998; Glazebrook
et al. 1999), albeit with very different sample selection
methods. Nevertheless, it is clear that it is now necesary to build
sizeable samples of H$\alpha$-selected star-forming galaxies at
different redshifts {\it and} study their properties. One would like
to know the preferred sites of star formation in the local universe
and beyond, and the main propeties of the star-forming galaxies and
their evolution.  Some questions that need to be answered include:
does star formation mainly occur in dwarf, starbursting galaxies or in
more quiescent, normal L$^*$ galaxies? how has that evolved with time? 
what fraction of the stellar mass of the galaxies is being built by
their current star-formation episodes?

Progress towards answering questions such as these requires, as a first
step, a comprehensive study of the properties of the star-forming
galaxies in the local universe. The Universidad Complutense de Madrid
survey (UCM hereafter; Zamorano et al. 1994, 1996) is currently the
most complete local sample of galaxies selected by their H$\alpha$
emission (see section~\ref{ucm}). It has been used to determine the
local H$\alpha$ luminosity function, the SFR function and the 
SFR density (Gallego et al.  1995). It is also widely used as a
benchmark for high redshift studies (e.g., Madau et al. 1998 and
references therein).  Thus, the UCM survey provides a suitable sample
of local star-forming galaxies for detailed studies.

Both optical imaging (Gunn-$r$; Vitores et al. 1996a, 1996b) and
spectroscopy of the whole UCM sample (Gallego et al. 1996; GAL96
hereafter; see also Gallego et al. 1997) are already available. The
optical data provides information on the current star-formation
activity, but is rather insensitive to the past star-formation history
of the galaxies.  In this paper we present new near-infrared imaging
observations for a representative subsample of UCM galaxies. The near
infrared luminosities are sensitive to the mass in older stars, and
therefore provide a measurement of the integrated past star formation
in the galaxies and their total stellar masses (see, e.g.,
Arag\'on-Salamanca et al. 1993; Alonso-Herrero et al. 1996; Charlot
1998).  Alonso-Herrero et al. (1996; AH96 hereafter) carried out a
pilot study of similar nature with a very small sample.  We will now
extend the work to a galaxy sample that is large enough for
statistical studies, and that is expected to represent the properties
of the complete UCM sample and thus those of the local star-forming
galaxy population. We will also improve the work of AH96 in two
fronts: first, we will use up-to-date population synthesis models, and
second, we will use a more sofisticated statistical technique when
comparing observational data and model predictions.

In section~\ref{ucm} we briefly introduce the UCM sample. In
section~\ref{observations} the observations, reduction procedures, and
data analysis are described. The evolutionary synthesis models are
presented in section~\ref{models}, and the results are described in
section~\ref{results}. Finally, section~\ref{summary} contains a
summary of this work.
  
\section{UCM survey} 
\label{ucm}

The UCM survey is a wide-field objective-prism search for star-forming
galaxies which used the H$\alpha$ emission line as main selection
criterium (Zamorano et al. 1994, 1996). This survey was carried out at
the 80-120cm Schmidt Telescope of the Calar Alto German-Spanish
Observatory (Almer\'{\i}a, Spain), using IIIaF photographic plates. The
identification of the emission-line objects was done by visual
inspection of the plates over the 471.4 square degrees that the survey
covers. An automatic procedure for the detection has also been
developed by Alonso et al. (1995, 1999) which avoids possible human
subjectivities in the selection.  The number of emission-line
candidates found was 264, about 44~per cent of them previously
uncatalogued. This yielded a detection rate of about 0.6 objects per
square degree \cite{jaz94}. A total of 191 of these objects were
confirmed spectroscopically by GAL96 as emission-line galaxies.

The wavelength cut-off of the photographic emulsion limits the
redshift range spanned by the survey to H$\alpha$-emitting objects
below $z$=0.045$\pm$0.005. The completeness tests performed (Vitores
1994; Gallego 1995) ensure that the Gunn-$r$ limiting magnitude of the
whole sample is 16.5$^{\mathrm{m}}$ with an H$\alpha$ equivalent width
detection limit of about 20\AA.

Details about the observations, data reduction, reliability and
accessibility of the complete data set are summarized in Zamorano et
al. (1994, 1996).

\section{Observational data} 
\label{observations}

\subsection{Optical imaging} 

The complete description of the optical Gunn-$r$ \cite{thuan76}
observations and image reduction is given in Vitores et al. (1996a,
1996b). Briefly, these images were acquired during a total of eight
observing runs from December 1988 through January 1992 using different
CCD detectors on the CAHA/MPIA 2.2-m and 3.5-m telescopes, both at Calar
Alto (Almer\'{\i}a, Spain).

\subsection{Near-infrared images} 

Near-infrared (nIR hereafter) images in the $J$ (1.2$\mu$m) and $K$
(2.2$\mu$m) or $K'$-bands (2.1$\mu$m), were obtained for 67 galaxies
from the UCM survey during three observing runs at the Lick Observatory
and one run at the Calar Alto Observatory.

The three Lick observing runs took place in 1996 (January 9--14, May
4--7 and June 7--9). We used the Lick InfraRed Camera (LIRC-II)
equipped with a {\sc nicmos3} 256$\times$256 detector on the 1-m
telescope of the Lick Observatory (California, USA). The instrumental
setup provided a total field of view of 2.4$\times$2.4 square arc
minutes with a spatial scale of 0.57\arcsec\ per pixel. For details
about the LIRC-II camera and the {\sc nicmos3} detector used see Misch,
Gilmore \& Rank \shortcite{misch95}. The Calar Alto observations were
carried out in 1996 (August 4--6). We used the MAGIC camera with a {\sc
nicmos3} 256$\times$256 detector attached to the CAHA/MPIA 2.2-m
telescope at Calar Alto (Almer\'{\i}a, Spain). The field of view was
2.70$\times$2.70 square arc minutes and the spatial scale
0.63\arcsec\ per pixel. Details about the MAGIC camera can be found in
Herbst et al. \shortcite{herbst93}. Images were obtained in the $J$ and
$K'$ bands in all the observing runs, except for the January 9--12 one,
when a standard $K$ filter was used. A $K'$ filter \cite{wainscoat}
was used in order to reduce the thermal background introduced by the
red wing of the standard $K$ passband.

The observational procedure followed was extensively described in
Arag\'on-Salamanca et al. \shortcite{alfonso93}. Briefly, we subdivided
the total exposure time required for each object in a number of images,
offset by a few arcseconds, in order to avoid saturation. Also, blank
sky images were obtained between consecutive object images for
sky-subtraction and flat-fielding purposes with offests larger than
$\sim1$ arc minute.  This procedure allows to reduce the effect of
pixel-to-pixel variations, bad pixels, cosmic rays, and faint star
images in the sky frames.

The reduction was carried out using our own IRAF\footnote{IRAF is
distributed by the National Optical Astronomy Observatories, which is
operated by the Association of Universities for Research in Astronomy,
Inc. (AURA) under cooperative agreement with the National Science
Foundation.} procedures, following standard reductions steps also
described in Arag\'on-Salamanca et al. \shortcite{alfonso93}, these
included bad pixel removal, dark substraction, flat-fielding, and sky
substraction.  Finally, all the object images were aligned, combined
and flux calibrated. Flux calibration was performed using standard
stars from the lists of Elias \shortcite{elias82} and Courteau
\shortcite{courteau95} observed at airmasses close to those of the
objects. The atmospheric extinction coefficients used were
$\kappa_{J}$=0.102\,mag/airmass and $\kappa_{K}$=0.09\,mag/airmass,
while independent zero-points were derived for each night.  The
$K'$-band magnitudes were converted into $K$-band magnitudes using the
empirical relation given by Wainscoat \& Cowie \shortcite{wainscoat},
$K'-K$=0.22$\times(H-K)$. Based on the zero-redshift SEDs of
Arag\'on-Salamanca et al. \shortcite{alfonso93} and the mean nIR
colours given in AH96 for a small sample of UCM galaxies, we used an
$H-K$ colour of 0.3$\pm$0.1$^{\mathrm{m}}$. Thus, the $K'-K$ correction
was 0.07$\pm$0.02$^{\mathrm{m}}$.

Aperture photometry was carried out on the Gunn-$r$ and nIR images
using the {\sc IRAF/APPHOT} routines. We measured $rJK$ magnitudes and
optical-nIR colours ($r-J$, $J-K$) through several physical apertures,
including three disk-scale
lengths\footnote{$H_0$=50\,km\,s$^{-1}$\,Mpc$^{-1}$ and $q_0$=0.5 have
been assumed throughout this paper} (see Vitores et al. 1996a). We
also measured total $K$-band magnitudes using physical apertures large
enough to ensure that all the light from the galaxy was included.
Finally, we corrected for contamination from field stars by replacing
affected pixels by adjacent sky counts. Circular aperture $r-J$ and
$J-K$ colours measured at three disk-scale lengths and integrated
$K$-band magnitudes are given in Table~\ref{data}. Magnitude and
colour errors include both calibration and photometric
uncertainties. In all the objects analyzed, except UCM0014$+$1748 and
UCM1432$+$2645 (which was observed off-centre), the field covered by
the detector was large enough to include three disk-scale length
apertures. For these two galaxies we obtained the three disk-scale
colours from their extrapolated growth curves. The differences between
the larger measurable aperture and the extrapolated values were
0.05$^{\mathrm{m}}$ for UCM0014$+$1748 and 0.1$^{\mathrm{m}}$ for
UCM1432$+$2645.

In our analysis, the $rJK$ magnitudes and optical-nIR colours are
corrected for Galactic and internal extinction. Since most of the
luminosity of these galaxies in the $rJK$ passbands comes from the
stellar continuum, we have estimated the colour excesses on the
continuum, $E(B-V)_{\mathrm{continuum}}$, with the expression given by
Calzetti, Kinney \& Storchi-Bergmann (1996, see also Calzetti 1997a;
Storchi-Bergmann, Calzetti \& Kinney 1994),
\begin{equation}
E(B-V)_{\mathrm{continuum}} = 0.44 \times E(B-V)_{\mathrm{gas}}.
\end{equation} 
where the $E(B-V)_{\mathrm{gas}}$ were obtained from the spectroscopic
Balmer decrements measured by GAL96 (see below). We assumed a diffuse
dust model that implies a total-to-selective extinction ratio of
$R_V$=3.1 (see Mathis 1990; Cardelli, Clayton \& Mathis
1989). Assuming a Galactic extinction curve, we obtained that
$A_r/A_V$, $A_J/A_V$ and $A_K/A_V$ are 0.83, 0.28 and 0.11
respectively \cite{mathis90}. Note than when correcting the H$\alpha$
fluxes and equivalent widths we assume that the line emission comes
from the gaseous component, but the continuum is mainly stellar. In
table~\ref{data} we present the observational data before correcting
for extinction, together with the gas colour excesses needed for the
correction.

\begin{table*}
\begin{minipage}{180mm}
\caption[]{Optical-nIR colours at three disk-scale length apertures
($d_{\mathrm{L}}$=disk-scale radius, as given by Vitores et al.,
1996a), and integrated $K$-band magnitudes.  Redshifts,
EW(H$\alpha$+[NII]), $L_{{\rm H}\alpha}$, E($B-V$)$_{\mathrm{gas}}$
and spectroscopic types have been taken from GAL96. In this table only
the H$\alpha$ luminosity data are given corrected for extinction.}
\begin{tabular}{lccccccccc}
Galaxy & Redshift & $d_{\mathrm{L}}$(kpc) &
$r-J$ & $J-K$ & $K$ & EW$^{\dag}$~(\AA) &  
$L_{\mathrm{H}\alpha}$ (10$^{8}$\,$L_{\sun}$) &
E($B-V$)$_{\mathrm{gas}}$ & Type \\
\hline
0003$+$2200  & 0.0245 &  1.66  &  1.59$\pm$0.28 &  1.05$\pm$0.44 & 12.83$\pm$0.35 &  50$\pm$1 & 0.28 & 0.87 & DANS\\ 
0013$+$1942  & 0.0270 &  2.02  &  1.54$\pm$0.08 &  0.95$\pm$0.10 & 14.11$\pm$0.07 & 142$\pm$4 & 0.86 & 0.28 & HIIH\\ 
0014$+$1748  & 0.0182 & 13.76  &  2.07$\pm$0.11 &  0.93$\pm$0.09 & 11.08$\pm$0.05 & 135$\pm$2 & 2.95 & 0.81 & SBN\\  
0014$+$1829  & 0.0182 &  0.88  &  1.32$\pm$0.18 &  0.95$\pm$0.24 & 12.96$\pm$0.20 & 146$\pm$2 & 0.56 & 1.47 & HIIH\\ 
0015$+$2212  & 0.0199 &  1.31  &  1.70$\pm$0.11 &  0.94$\pm$0.10 & 13.21$\pm$0.07 & 147$\pm$2 & 1.01 & 0.22 & HIIH\\ 
0017$+$1942  & 0.0259 &  2.87  &  1.38$\pm$0.13 &  0.84$\pm$0.11 & 13.11$\pm$0.07 & 181$\pm$6 & 2.96 & 0.36 & HIIH\\ 
0022$+$2049  & 0.0185 &  1.57  &  2.15$\pm$0.12 &  1.14$\pm$0.10 & 11.19$\pm$0.05 & 106$\pm$2 & 1.83 & 0.90 & HIIH\\ 
0050$+$2114  & 0.0245 &  2.39  &  2.00$\pm$0.12 &  0.99$\pm$0.13 & 11.65$\pm$0.09 & 111$\pm$1 & 2.76 & 0.81 & SBN\\  
0145$+$2519  & 0.0409 &  5.81  &  1.77$\pm$0.13 &  1.10$\pm$0.14 & 12.11$\pm$0.10 &  38$\pm$1 & 2.60 & 1.03 & SBN\\  
1255$+$3125  & 0.0252 &  1.89  &  1.90$\pm$0.15 &  0.88$\pm$0.23 & 12.50$\pm$0.18 &  74$\pm$1 & 1.45 & 0.41 & HIIH\\ 
1256$+$2823  & 0.0315 &  2.58  &  1.56$\pm$0.11 &  1.15$\pm$0.15 & 12.46$\pm$0.11 & 109$\pm$2 & 2.82 & 0.64 & SBN\\  
1257$+$2808  & 0.0171 &  1.11  &  1.35$\pm$0.33 &  1.35$\pm$0.44 & 12.78$\pm$0.29 &  42$\pm$1 & 0.29 & 1.34 & SBN\\  
1259$+$2755  & 0.0240 &  2.65  &  1.42$\pm$0.17 &  1.10$\pm$0.18 & 11.89$\pm$0.13 &  62$\pm$1 & 1.81 & 0.91 & SBN\\  
1259$+$3011  & 0.0307 &  2.01  &  1.45$\pm$0.15 &  1.28$\pm$0.19 & 12.63$\pm$0.14 &  34$\pm$1 & 0.75 & 0.68 & SBN\\  
1302$+$2853  & 0.0237 &  1.55  &  1.73$\pm$0.15 &  0.92$\pm$0.25 & 12.85$\pm$0.20 &  48$\pm$1 & 0.43 & 0.62 & DHIIH\\
1304$+$2808  & 0.0205 &  2.84  &  1.63$\pm$0.14 &  1.26$\pm$0.19 & 12.02$\pm$0.14 &  33$\pm$1 & 0.55 & 0.11 & DANS\\ 
1304$+$2818  & 0.0243 &  2.82  &  1.43$\pm$0.11 &  1.12$\pm$0.12 & 12.42$\pm$0.10 &115$\pm$11 & 2.31 & 0.11 & SBN\\ 
1306$+$2938  & 0.0209 &  1.64  &  1.45$\pm$0.10 &  1.15$\pm$0.11 & 12.15$\pm$0.09 & 133$\pm$4 & 2.11 & 0.50 & SBN\\  
1307$+$2910  & 0.0187 &  5.81  &  1.71$\pm$0.34 &  1.21$\pm$0.43 & 10.37$\pm$0.28 &  39$\pm$1 & 2.56 & 0.97 & SBN\\  
1308$+$2950  & 0.0242 &  7.58  &  2.10$\pm$0.10 &  1.17$\pm$0.14 & 10.75$\pm$0.10 &  59$\pm$1 & 3.08 & 1.38 & SBN\\  
1308$+$2958  & 0.0212 &  4.17  &  1.71$\pm$0.06 &  0.77$\pm$0.16 & 12.03$\pm$0.15 &  26$\pm$1 & 7.03 & 1.31 & SBN\\  
1312$+$2954  & 0.0230 &  2.97  &  1.90$\pm$0.14 &  1.06$\pm$0.35 & 12.15$\pm$0.32 &  65$\pm$3 & 0.96 & 1.09 & SBN\\  
1312$+$3040  & 0.0210 &  2.31  &  1.83$\pm$0.10 &  1.13$\pm$0.09 & 11.69$\pm$0.07 &  81$\pm$2 & 1.51 & 0.47 & SBN\\  
1428$+$2727  & 0.0149 &  1.48  &  0.81$\pm$0.18 &  0.84$\pm$0.22 & 12.45$\pm$0.17 & 218$\pm$3 & 2.35 & 0.15 & HIIH\\ 
1432$+$2645  & 0.0307 &  6.47  &  1.71$\pm$0.10 &  1.15$\pm$0.14 & 11.84$\pm$0.10 &  47$\pm$1 & 2.09 & 0.91 & SBN\\  
1440$+$2511  & 0.0333 &  4.40  &  1.71$\pm$0.05 &  1.32$\pm$0.13 & 12.89$\pm$0.12 &  35$\pm$1 & 0.57 & 1.02 & SBN\\  
1440$+$2521N & 0.0315 &  2.55  &  1.86$\pm$0.32 &  1.31$\pm$0.43 & 12.53$\pm$0.29 & 104$\pm$3 & 1.59 & 0.77 & SBN\\  
1440$+$2521S & 0.0314 &  2.06  &  1.40$\pm$0.33 &  1.46$\pm$0.45 & 13.25$\pm$0.30 & 100$\pm$5 & 1.04 & 0.29 & SBN\\  
1442$+$2845  & 0.0110 &  1.27  &  1.96$\pm$0.10 &  0.98$\pm$0.14 & 11.68$\pm$0.10 & 135$\pm$3 & 0.66 & 0.68 & SBN\\  
1443$+$2548  & 0.0351 &  3.05  &  2.00$\pm$0.36 &  0.61$\pm$0.44 & 12.59$\pm$0.26 &  76$\pm$1 & 2.63 & 0.73 & SBN\\  
1452$+$2754  & 0.0339 &  2.74  &  2.40$\pm$0.37 &  0.88$\pm$0.44 & 12.13$\pm$0.25 & 135$\pm$2 & 3.09 & 0.73 & SBN\\  
1506$+$1922  & 0.0205 &  3.00  &  2.12$\pm$0.36 &  1.00$\pm$0.44 & 11.78$\pm$0.25 & 140$\pm$6 & 1.95 & 0.45 & HIIH\\ 
1513$+$2012  & 0.0369 &  2.20  &  1.76$\pm$0.09 &  1.49$\pm$0.08 & 11.87$\pm$0.07 & 150$\pm$2 & 6.18 & 0.54 & SBN\\  
1557$+$1423  & 0.0275 &  1.85  &  1.82$\pm$0.11 &  1.01$\pm$0.09 & 12.92$\pm$0.06 &  54$\pm$1 & 0.64 & 0.37 & SBN\\  
1646$+$2725  & 0.0339 &  1.81  &  1.72$\pm$0.23 &  0.93$\pm$0.17 & 15.04$\pm$0.13 & 225$\pm$3 & 0.50 & 0.29 & DHIIH\\
1647$+$2729  & 0.0366 &  3.16  &  1.94$\pm$0.10 &  0.99$\pm$0.10 & 12.42$\pm$0.06 &  59$\pm$1 & 2.06 & 0.89 & SBN\\  
1647$+$2950  & 0.0290 &  2.97  &  1.88$\pm$0.34 &  0.91$\pm$0.43 & 11.91$\pm$0.28 & 110$\pm$2 & 3.76 & 0.74 & SBN\\  
1648$+$2855  & 0.0308 &  1.95  &  1.25$\pm$0.08 &  1.06$\pm$0.12 & 12.67$\pm$0.12 & 240$\pm$15& 6.23 & 0.25 & HIIH\\
1654$+$2812  & 0.0348 &  2.57  &  1.55$\pm$0.14 &  0.97$\pm$0.22 & 14.93$\pm$0.18 &  70$\pm$3 & 0.33 & 0.31 & DHIIH\\
1656$+$2744  & 0.0330 &  1.04  &  1.95$\pm$0.17 &  1.25$\pm$0.14 & 13.08$\pm$0.09 & 108$\pm$1 & 1.08 & 0.58 & SBN\\  
1657$+$2901  & 0.0317 &  1.33  &  1.60$\pm$0.13 &  1.29$\pm$0.15 & 13.37$\pm$0.12 &  80$\pm$1 & 0.68 & 0.56 & DANS\\ 
\hline
\end{tabular}
Note: $^{\dag}$ Equivalent width of H$\alpha$+[NII]
\end{minipage}
\end{table*}

\begin{table*}
\begin{minipage}{180mm}
\addtocounter{table}{-1}
\caption[]{ (cont.)}
\begin{tabular}{lccccccccc}
Galaxy & Redshift & $d_{\mathrm{L}}$(kpc) &
$r-J$ & $J-K$ & $K$ & EW$^{\dag}$~(\AA) &  
$L_{\mathrm{H}\alpha}$ (10$^{8}$\,$L_{\sun}$) &
E($B-V$)$_{\mathrm{gas}}$ & Type \\
\hline
2238$+$2308 & 0.0238 &  4.44  &  1.96$\pm$0.09 &  0.97$\pm$0.09 & 11.10$\pm$0.05 &  69$\pm$1 & 3.15 & 1.05 & SBN\\  
2239$+$1959 & 0.0242 &  2.79  &  1.62$\pm$0.10 &  1.03$\pm$0.08 & 11.52$\pm$0.04 & 173$\pm$5 & 5.00 & 0.54 & HIIH\\ 
2250$+$2427 & 0.0421 &  6.01  &  1.82$\pm$0.10 &  1.20$\pm$0.08 & 11.72$\pm$0.04 & 175$\pm$4 & 11.12& 0.71 & SBN\\  
2251$+$2352 & 0.0267 &  1.13  &  1.53$\pm$0.11 &  0.97$\pm$0.08 & 13.27$\pm$0.05 &  84$\pm$1 & 0.97 & 0.18 & DANS\\ 
2253$+$2219 & 0.0242 &  1.45  &  1.99$\pm$0.08 &  1.10$\pm$0.08 & 12.40$\pm$0.05 &  86$\pm$1 & 1.07 & 0.54 & SBN\\  
2255$+$1654 & 0.0388 &  5.75  &  2.43$\pm$0.09 &  1.37$\pm$0.10 & 11.56$\pm$0.05 &  40$\pm$1 & 1.39 & 1.47 & SBN\\  
2255$+$1926 & 0.0193 &  2.13  &  1.47$\pm$0.13 &  0.96$\pm$0.11 & 13.56$\pm$0.07 &  37$\pm$1 & 0.16 & 0.37 & DHIIH\\
2255$+$1930N& 0.0189 &  2.11  &  2.00$\pm$0.10 &  1.07$\pm$0.09 & 11.63$\pm$0.05 &  97$\pm$1 & 1.38 & 0.70 & SBN\\  
2255$+$1930S& 0.0189 &  1.12  &  1.86$\pm$0.10 &  0.99$\pm$0.09 & 12.74$\pm$0.05 &  61$\pm$1 & 0.52 & 0.49 & SBN\\  
2258$+$1920 & 0.0220 &  1.87  &  2.04$\pm$0.10 &  0.98$\pm$0.10 & 12.35$\pm$0.06 & 190$\pm$2 & 1.75 & 0.35 & DANS\\ 
2300$+$2015 & 0.0346 &  2.42  &  2.05$\pm$0.12 &  1.10$\pm$0.10 & 12.70$\pm$0.05 & 159$\pm$1 & 3.17 & 0.33 & SBN\\  
2302$+$2053E& 0.0328 &  3.09  &  1.76$\pm$0.12 &  1.10$\pm$0.10 & 11.55$\pm$0.05 &  36$\pm$1 & 1.68 & 1.30 & SBN\\ 
2302$+$2053W& 0.0328 &  2.00  &  1.73$\pm$0.12 &  0.97$\pm$0.11 & 14.26$\pm$0.08 & 260$\pm$2 & 1.38 & 0.46 & HIIH\\  
2303$+$1856 & 0.0276 &  2.35  &  2.10$\pm$0.13 &  1.16$\pm$0.10 & 11.35$\pm$0.05 &  79$\pm$1 & 2.41 & 1.20 & SBN\\  
2304$+$1640 & 0.0179 &  1.00  &  1.30$\pm$0.15 &  0.93$\pm$0.19 & 14.82$\pm$0.15 & 155$\pm$2 & 0.20 & 0.33 & BCD \\ 
2307$+$1947 & 0.0271 &  1.63  &  2.10$\pm$0.22 &  1.10$\pm$0.14 & 12.41$\pm$0.08 &  45$\pm$1 & 0.64 & 0.45 & DANS\\ 
2310$+$1800 & 0.0363 &  2.84  &  2.31$\pm$0.14 &  1.16$\pm$0.14 & 12.26$\pm$0.08 &  63$\pm$1 & 1.46 & 0.90 & SBN\\  
2313$+$1841 & 0.0300 &  2.42  &  1.99$\pm$0.12 &  0.97$\pm$0.14 & 13.03$\pm$0.09 &  82$\pm$2 & 0.72 & 0.91 & SBN\\  
2316$+$2028 & 0.0263 &  1.42  &  2.75$\pm$0.14 &  1.03$\pm$0.14 & 12.81$\pm$0.09 &  99$\pm$4 & 0.49 & 0.75 & DANS\\ 
2316$+$2457 & 0.0277 &  3.77  &  2.01$\pm$0.12 &  1.14$\pm$0.14 & 10.41$\pm$0.08 & 109$\pm$1 & 10.58& 0.69 & SBN\\  
2316$+$2459 & 0.0274 &  4.08  &  2.17$\pm$0.13 &  0.94$\pm$0.14 & 11.90$\pm$0.08 &  72$\pm$1 & 10.16& 0.57 & SBN\\  
2319$+$2234 & 0.0364 &  2.69  &  2.46$\pm$0.12 &  1.05$\pm$0.14 & 12.98$\pm$0.09 & 108$\pm$1 & 1.91 & 0.59 & SBN\\  
2321$+$2149 & 0.0374 &  2.74  &  1.70$\pm$0.13 &  0.92$\pm$0.14 & 13.27$\pm$0.09 &  68$\pm$1 & 1.38 & 0.56 & DANS\\ 
2324$+$2448 & 0.0123 &  3.14  &  2.21$\pm$0.12 &  0.99$\pm$0.14 &  9.60$\pm$0.08 &  10$\pm$1 & 1.20 & 0.26 & SBN\\  
2327$+$2515N& 0.0206 &  1.39  &  1.20$\pm$0.12 &  0.92$\pm$0.21 & 13.10$\pm$0.18 & 289$\pm$7 & 1.83 & 0.47 & HIIH\\ 
2327$+$2515S& 0.0206 &  1.80  &  1.22$\pm$0.12 &  0.92$\pm$0.19 & 12.82$\pm$0.15 & 104$\pm$1 & 1.07 & 0.36 & HIIH\\ 
\hline
\label{data}
\end{tabular}
Note: $^{\dag}$ Equivalent width of H$\alpha$+[NII]
\end{minipage}
\end{table*}

\subsection{Optical spectroscopy} 

Optical long-slit spectroscopy for the UCM survey was obtained by
Gallego \shortcite{gallego95a} at the 2.5-m Isaac Newton Telescope
(INT) at Roque de los Muchachos Observatory, La Palma (Spain), and the
2.2-m and 3.5-m telescopes at Calar Alto (Almer\'{\i}a, Spain), during
a total of 10 observing runs. Details about the instrumental setups,
slit widths, spatial scales and dispersions achieved are given in
Table~1 of GAL96.

The line fluxes and equivalent widths of different emission lines are given in
GAL96. Gas colour excesses, $E(B-V)_{\mathrm{gas}}$, were obtained assuming a
Galactic extinction curve and intrinsic ratios
I$(\mathrm{H}\alpha)/I(\mathrm{H}\beta)$=2.86 and
I$(\mathrm{H}\gamma)/I(\mathrm{H}\beta)$=0.468, which are the theoretical
values expected for a low density gas with $T_{\mathrm{e}}$=10$^{4}$\,K in
Case~B recombination \cite{osterbrook89}. We estimate that the effect of
differential atmospheric  refraction on the H$\alpha/$H$\beta$ ratio is in most
cases (82\% of the galaxies) below 5\%. Only in four of the galaxies studied
here the  uncertainty in $E(B-V)_{\mathrm{gas}}$ due to differential 
refraction is larger than 0.1$^{\mathrm{m}}$.  The observation and
analysis procedures followed by Gallego \shortcite{gallego95a} ---slit widths,
position angles, spectrum extraction--- ensure good integrated spectroscopic
information for these objects. Errors in the EW(H$\alpha$+[NII]) have been
estimated from the signal-to-noise and spectral resolution data given by GAL96.
We have assumed a 100\AA\ interval for the continuum fit range,
$\Delta\lambda_{\mathrm{cont}}$ \cite{gallego95a}, and a reciprocal dispersion
of $\Delta\lambda\!\sim$3\AA/pixel. Thus, \begin{equation} {\Delta\mathrm{EW}}
= {1 \over {\mathrm{SNR}\sqrt{\mathrm{N}}}} \sqrt{\mathrm{EW}^2 +
\mathrm{FWZI}^2} \end{equation}  where SNR is the signal-to-noise ratio of the
continuum, N is the number of points used to determine the mean continuum flux,
i.e. N=$\Delta\lambda_{\mathrm{cont}}/\Delta\lambda\!\sim$30, and FWZI is the
{\it Full Width at Zero Intensity}. The FWZI was computed as two times the {\it
Full Width at Half Maximum} (FWHM) of the comparison arc lines. Typical FWHMs
are about 12.5\AA\ (for a 3\arcsec\ wide slit with the R300V$+$IDS$+$Tek3
configuration). Equivalent widths (EWs hereafter) of H$\alpha$+[NII] and their
corresponding errors are given in Table~\ref{data}. The H$\alpha$ equivalent
width data used in this work were corrected for contamination of the
[NII]$\lambda$6548\AA\ and [NII]$\lambda$6584\AA\ emission lines using the
[NII]$\lambda$$\lambda$6548,6584\AA/H$\alpha$ line ratios given by GAL96 (see
also Gallego 1995). Finally, we consider a correction of 2\,\AA\ in the
EW(H$\alpha$) due to the H$\alpha$ underlying absorption in G-K giants (see,
e.g., Kennicutt 1983).

\section{Evolutionary synthesis models}
\label{models}

Although the $K$-band luminosity can provide a very good estimate of
the stellar mass in galaxies (Arag\'on-Salamanca et al. 1993; Charlot
1998), the contribution from red supergiants associated with recent
star forming events may lead to the overestimate of the stellar mass
when standard mass-luminosity relations are used.  Thus, in order to
estimate the relative contribution of the old underlying and young
stellar populations to the magnitudes and colours measured, we have
developed a complete set of evolutionary synthesis models. These
models are based on those developed by AH96, but use the new
population synthesis models of Bruzual \& Charlot (private
communication, BC96 hereafter), instead of the old Bruzual \& Charlot
(1993; BC93 hereafter) models. From the number of ionizing photons
supplied by the BC96 models, we have also calculated the contribution
of the hydrogen and helium emission-lines and nebular continuum to the
optical and nIR passbands.

AH96 demonstrated that the properties of most of the UCM star-forming
galaxies are better reproduced with instantaneous burst models rather
that models with constant star formation. Therefore, we have computed
the evolution with time of the optical-nIR colours and EW(H$\alpha$)
of an instantaneous burst superimposed on a 15\,Gyr old evolving
population. A Scalo Initial Mass Function (IMF; Scalo 1986) with lower
and upper mass cutoffs of $M_{\mathrm{low}}$=0.1\,M$_{\sun}$ and
$M_{\mathrm{up}}$=125\,M$_{\sun}$ was adopted. The Cousins-$R$
magnitudes given by the BC96 models have been converted into Gunn-$r$
magnitudes using the relation
$r$=$R_{\mathrm{C}}+$0.383$-$0.083$\times(V-R_{\mathrm{C}})$ (Fernie
1983; Kent 1985).

In order to match the colours predicted by the BC96 models for a
15\,Gyr old Single Stellar Population (SSP hereafter; $r-J$=2.09,
$J-K$=0.85) to those measured in the bulges of local relaxed spiral
galaxies, assuming a negligible dust reddening (see Peletier \&
Balcells 1996; Fioc \& Rocca-Volmerange 1999), we applied a small
correction to our models \\ $(r-J)^{\mathrm{obs}}$=$(r-J)_{15
\mathrm{Gyr}}^{\mathrm{mod}}-0.03$\\
$(J-K)^{\mathrm{obs}}$=$(J-K)_{15\mathrm{Gyr}}^{\mathrm{mod}}+0.06$.\\
In addition, the stellar mass-to-light ratio predicted by the model
for a 15\,Gyr old stellar population, 1.34\,M$_{\sun}$/L$_{K,\sun}$,
was corrected to match that measured in local relaxed spiral galaxies,
$\sim$1\,M$_{\sun}$/L$_{K,\sun}$ (see H\'eraudeau \& Simien 1997, and
references therein), using M$_{K,\sun}$=3.33 \cite{worthey}. Since
most UCM galaxies (about 83~per cent) are morphologically classified
as Sa--Sc$+$ \cite{alvaro94}, we are confident that the corrections
applied to the models are reasonable. In any case, these small
corrections, intended to provide a good agreement between the model
predictions and observations for the underlying stellar populations of
the galaxies, do not affect significantly any of the conclusions of
this paper.

The main parameters of our models are those inherent to the BC96
models (age, metallicity, IMF, etc), together with the strength of the
current star-forming burst. The burst strength, $b$, is defined as the
ratio of the mass of the newly formed stars to the total stellar mass
of the galaxy \cite{kruger}. We have explored models with
metallicities between 1/50\,Z$_{\sun}$ and 2\,Z$_{\sun}$, and burst
strengths in the range 1--$10^{-4}$ (in 0.04\,dex steps). The new BC96
models (Scalo IMF) produce slightly redder $r-J$ and $J-K$ colours and
fewer Lyman photons than the BC93 ones (Salpeter IMF), for the same
burst strength and solar metallicity.

\section{Results}
\label{results}

The main goal of this work is the characterization of the star
formation activity of a representative sample of local galaxies.  The
properties of the current star-formation events and the host galaxies
will be studied. In particular, we are interested in linking the
properties of the local star-forming  galaxies with those of galaxies
forming stars at higher redshifts.

First, we study the completeness and representativeness of our sample
in relation to the local star-forming galaxy population (see
section~\ref{completeness}). In section~\ref{colours} we analyze the
measured magnitudes and colours of the galaxies.  Then, and in order to
obtain the burst strengths, burst ages, stellar masses and {\it
specific\/} star formation rates (SFR per unit mass; Guzm\'an et al.
1997; Lowenthal et al. 1997), we compare our data with evolutionary
synthesis models (section~\ref{minimization}). The comparison between
data and models, and the determination of the best-fitting set of
parameters, are not straightforward tasks, and some details of the
method we have applied are described in the Appendix. Finally, we
discuss the derived burst strengths, ages, metallicities, galaxy
stellar masses and star formation rates derived for our sample
(section~\ref{strength}, \ref{mass} and \ref{sfr}).

\begin{figure}
\psfig{bbllx=125.,bblly=120.,bburx=400.,bbury=835.,file=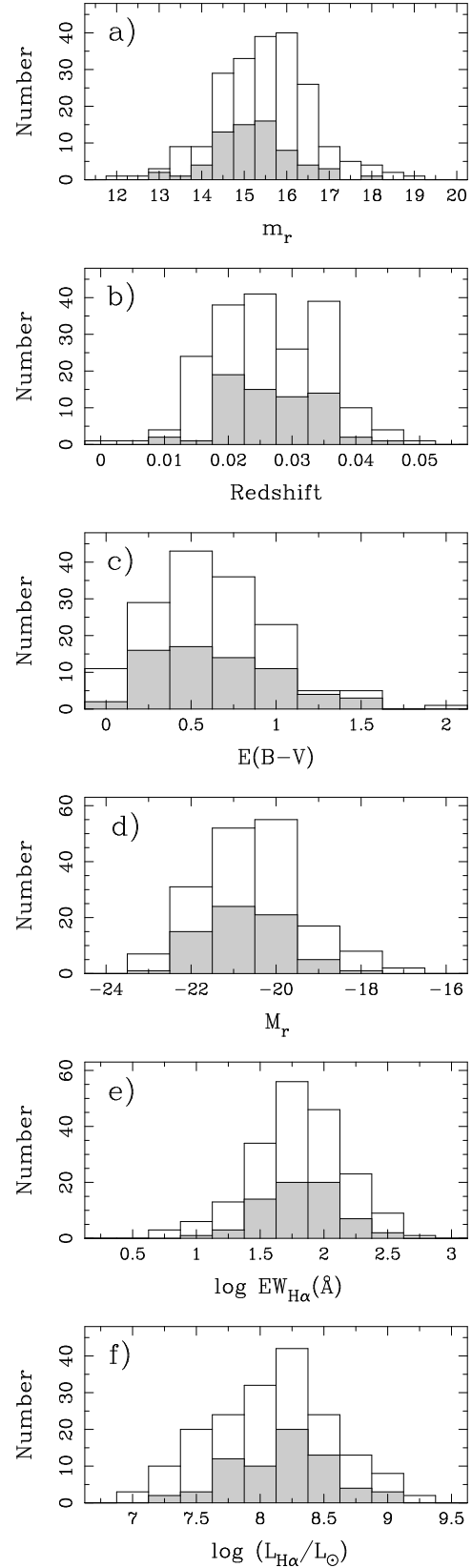,height=22.0cm,width=8.2cm,angle=0}
\caption{From top to bottom, distributions of the observed 
Gunn-$r$ magnitudes, redshifts, gas colour excesses, absolute Gunn-$r$
magnitudes, EW(H$\alpha$), and H$\alpha$ luminosities. The open histograms 
correspond to the complete UCM sample, and the grey-filled areas correspond to the nIR sample.}
\label{histdata}
\end{figure}

\subsection{Sample completeness}
\label{completeness}

Our nIR sample will suffer, first, from the intrinsic selection
effects of the objective-prism$+$photographic plate technique used in
the UCM survey. Those were discussed in detail in Zamorano et
al. (1994, 1996), Vitores \shortcite{alvaro94} and Gallego
\shortcite{gallego95a}. Briefly, the observational procedure employed 
limits the UCM sample to local galaxies with redshift lower than
0.045$\pm$0.005 and H$\alpha$ equivalent width larger than 20\AA. The
Gunn-$r$ limiting magnitude is 16.5$^{\mathrm{m}}$ with a brigth end
cut-off, due to saturation of the photographic plates, placed at
$\sim$14.2$^{\mathrm{m}}$. However, additional selection effects may
be present in our work due to the limited size of our nIR sample
($\sim$35 per cent of the UCM whole sample), thus it is necessary to
ensure that the properties of this subsample are representative of
those of the complete UCM survey.

In Figure~\ref{histdata}a we compare the Gunn-$r$ apparent magnitude
histogram of the whole UCM sample with that of the galaxies observed in
the nIR. Although the apparent magnitude distributions match reasonably
well, the objects in the nIR subsample tend to be marginally brighter
than those in the UCM complete sample. The median $m_r$ for the UCM
complete sample is 15.5$^{\mathrm{m}}$, while that of the nIR sample is
15.2$^{\mathrm{m}}$ (see Vitores et al. 1996b). This may imply a small
deficiency of low-luminosity and/or higher redshift galaxies. However,
that does not seem to be a strong effect (see figures~\ref{histdata}b
and~d).

From Figures~\ref{histdata}b, \ref{histdata}c and \ref{histdata}e it
is clear that the nIR sample represents about 35~per cent of the UCM
complete sample \cite{gallego95a} in every redshift, $E(B-V)$ and
EW(H$\alpha$) bin. A Kolmogorov-Smirnov test indicates that both
samples show similar distributions in $z$, $E(B-V)$ and EW(H$\alpha$),
with probabilities of 45, 93 and 87~per cent, respectively. For the
$r$-band absolute magnitudes and H$\alpha$ luminosities the
probabilities are 25 and 47~per cent, respectively (see
Figures~\ref{histdata}d and \ref{histdata}f). Thus the galaxies in the
nIR sample seem to be a fair subsample of the UCM complete sample in
their global properties.

The only small difference arises when comparing in detail the
spectroscopic type distributions of the star-forming galaxies in the
nIR and UCM complete samples.  There is a small deficiency of HII-{\it
like\/} galaxies (see ahead) in the nIR subsample relative to the
whole UCM sample. About 30~per cent of the UCM whole sample are
HII-{\it like\/} galaxies, whereas only 19~per cent are present in the
nIR subsample. Consequences of such a limitation will be taken into
account in further discussions.

\subsection{Aperture magnitudes and colours}
\label{colours}

\subsubsection{Mean colours}
\label{mean}

Global colours (obtained inside three disk-scale lengths), together
with the integrated $K$-band magnitudes (not corrected for extinction)
and the $E(B-V)_{\mathrm{gas}}$ colour excesses, are given in
Table~\ref{data}. In GAL96, the galaxies in the UCM sample were
classified in different morphological and spectroscopic classes
(listed in table~\ref{data}). We will briefly describe them 
here (see GAL96 for details):

\noindent
{\bf\Large SBN} ---{\it Starburst Nuclei}--- Originally defined by
Balzano \shortcite{balzano}, they show high extinction values, with
very low [NII]/H$\alpha$ ratios and faint [OIII]$\lambda$5007
emission. Their H$\alpha$ luminosities are always higher than
10$^{8}$\,L$_{\sun}$.

\noindent 
{\bf\Large DANS} ---{\it Dwarf Amorphous Nuclear Starburst}---
Introduced by Salzer, MacAlpine \& Boroson \shortcite{salzer}, they
show very similar spectroscopic properties to SBN objects, but with
H$\alpha$ luminosities lower than 5$\times$10$^{7}$\,L$_{\sun}$.

\noindent
{\bf\Large HIIH} ---{\it HII Hotspot}--- The HII Hotspot class shows (see
GAL96) similar H$\alpha$ luminosities to those measured in SBN
galaxies but with large [OIII]$\lambda$5007/H$\beta$ ratios, that is,
higher ionization.

\noindent
{\bf\Large DHIIH} ---{\it Dwarf HII Hotspot}---- This is an HIIH subclass
with identical spectroscopic properties but H$\alpha$ luminosities
lower than 5$\times$10$^{7}$\,L$_{\sun}$.

\noindent
{\bf\Large BCD} ---{\it Blue Compact Dwarf}--- Finally, the lowest
luminosity and highest ionization objects have been classified as Blue
Compact Dwarf galaxies, showing in all cases H$\alpha$ luminosities
lower than 5$\times$10$^{7}$\,L$_{\sun}$. They also show large
[OIII]$\lambda$5007/H$\beta$ and H$\alpha$/[NII]$\lambda$6584 line
ratios and intense [OII]$\lambda$3727 emission.

In our analysis, we separate the galaxies in two main categories:
starburst {\it disk-like} (SB hereafter) and HII-{\it like} galaxies
(see Guzm\'an et al. 1997; Gallego 1998). The SB-{\it like} class
includes SBN and DANS spectroscopic types, whereas the HII-{\it like}
includes HIIH, DHIIH and BCD type galaxies.

In order to determine representative mean optical-nIR colours for each
galaxy group, we have assumed Gaussian probability distributions for
the $r-J$ and $J-K$ colours and EW(H$\alpha$) with the centres and
widths ($\sigma$) given in Table~\ref{data}. We have weighted the data
for each galaxy with their corresponding errors when determining the
mean values.

The HII-{\it like} objects seem to be on average 0.2$^{\mathrm{m}}$
bluer in $r-J$ and 0.1$^{\mathrm{m}}$ in $J-K$ than the SB galaxies
(see Table~\ref{datamean}). Since the mean colour excess of the SB
population ($\overline{E(B-V)}$=0.7$^{\mathrm{m}}$) is
0.2$^{\mathrm{m}}$ higher than that of the HII-{\it like} galaxies,
these colour differences are even more significant when data not
corrected for extinction are used: the differences for the
un-corrected colours are 0.35$^{\mathrm{m}}$ in $r-J$ and
0.15$^{\mathrm{m}}$ in $J-K$. K-S tests performed on the SB-{\it like}
and HII-{\it like} objects indicate that both subsamples arise from
independent distributions with a probability of 99.7 and 99.9~per cent
respectively for the un-corrected $r-J$ and $J-K$ colours.

Finally, whereas more than 60~per cent of the HII-{\it like} objects
show not corrected for extinction equivalent widths of H$\alpha$
higher than 120\,\AA, only 3~per cent of the SB galaxies do. The
relatively low EW(H$\alpha$) detection limit estimated for the UCM
survey ($\sim$\,20\,\AA; Gallego 1995) ensures that the difference in
EW(H$\alpha$) between SB and HII-{\it like} galaxies is not due to
selection effects. A K-S test gives a probability of 99.9~per cent
that these samples have independent EW(H$\alpha$) distributions. These
differences in colours and H$\alpha$ equivalent widths are probably
related to differences in their evolutionary properties (typical
starburst age, starburst strength, {\it specific} star formation rate,
etc) between both galaxy types (see section~\ref{strength} and
section~\ref{sfr}).

\subsubsection{Colour-colour and colour-EW(H$\alpha$) diagrams}
\label{colcol}

\begin{figure*}
\psfig{bbllx=20.,bblly=310.,bburx=565.,bbury=785.,file=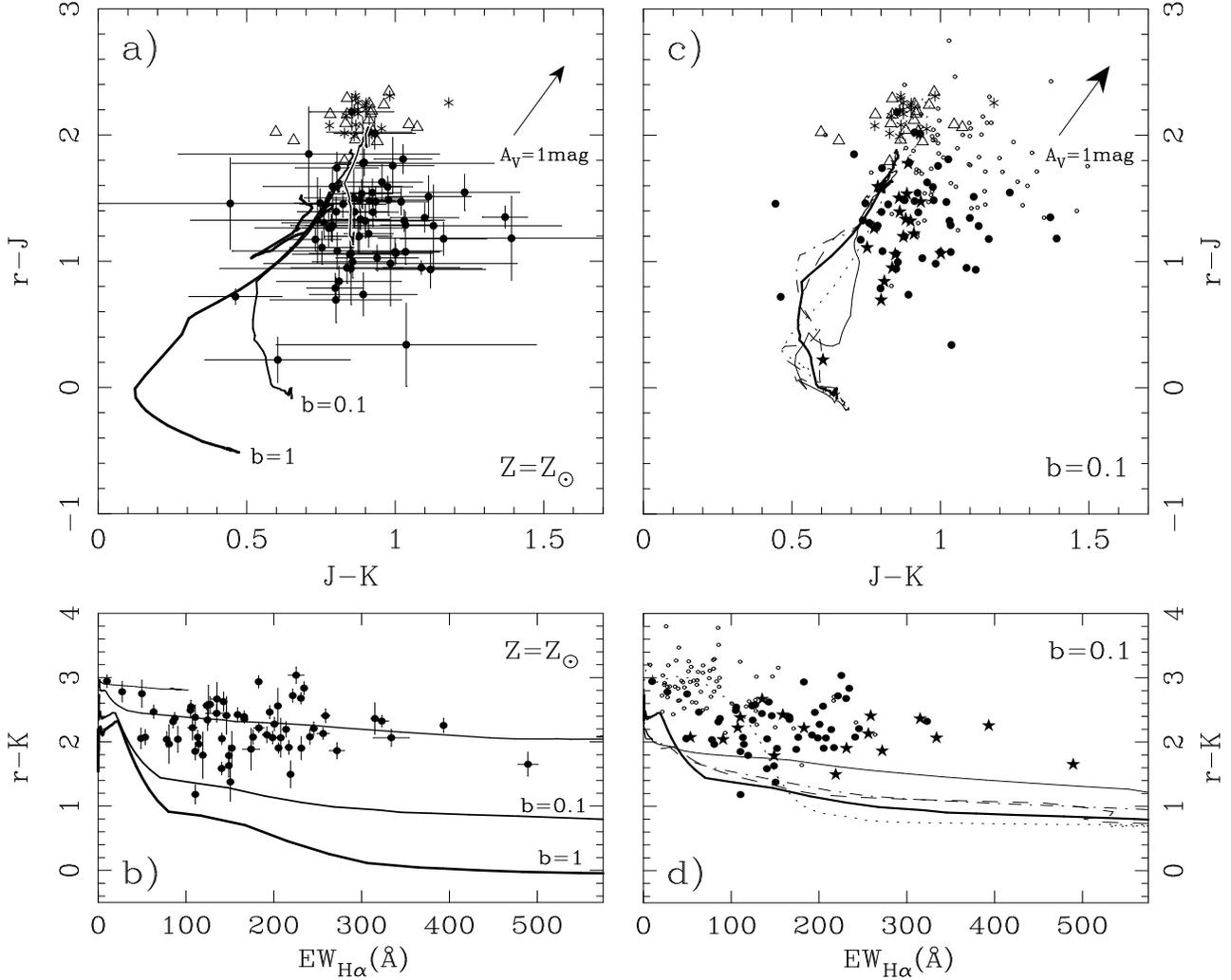,height=14cm,width=18cm,angle=0}
\caption{Colour-colour and colour-EW(H$\alpha$) diagrams. In
 panels {\bf a} and {\bf b} optical-nIR colours are plotted for the 67
 galaxies of the sample. Solar metallicity models have been plotted
 using progresively thicker lines for higher burst strenght models
 ($b$=10$^{-3}$,10$^{-2}$,10$^{-1}$,1). Data for nearby relaxed spiral
 galaxies have been taken from Peletier \& Balcells (1996); both bulge
 ({\it asterisks}) and disk ({\it triangles}) colours are
 shown. Panels {\bf c, d} show the optical-nIR colours of the SB-{\it
 like} ({\it dots}) and the HII-{\it like} ({\it stars}) objects. 
 Values not
 corrected for extinction are also shown ({\it small
 circles}). In panels {\bf c} \& {\bf d}, models with 10$^{-1}$ burst
 strength and different metallicities from 1/50\,Z$_{\sun}$ to
 2\,Z$_{\sun}$ have been drawn (1/50\,Z$_{\sun}$, {\it thin solid
 line}; 1/5\,Z$_{\sun}$, {\it dashed line}; 2/5\,Z$_{\sun}$, {\it
 dashed-dotted line}; Z$_{\sun}$, {\it thick solid line};
 2\,Z$_{\sun}$, {\it dotted line}). $\pm$1$\sigma$ error bars are also
 shown.}
\label{CCD}
\end{figure*}

In Figure~\ref{CCD} we show colour-colour ($r-J$ vs. $J-K$) and
colour-EW(H$\alpha$) plots for the nIR sample. The offset 
between the position of the star forming galaxies ({\it filled
circles}) in the $r-J$--$J-K$ plane (Figure~\ref{CCD}a) and the
bulges and disks of relaxed nearby spirals \cite{peletier96} indicates
the existence of ongoing star formation. Error bars in
Figure~\ref{CCD}a represent $\pm$1$\sigma$ errors. In
Figures~\ref{CCD}a and \ref{CCD}b we plot solar metallicity models
with burst strengths, 10$^{-3}$, 10$^{-2}$, 10$^{-1}$, and 1. In the
case of Figures~\ref{CCD}c and \ref{CCD}d, models with 10$^{-1}$ burst
strength and different metallicities are displayed (cf. section~\ref{models}).

Figures~\ref{CCD}a and \ref{CCD}c show that changes in the optical-nIR
colours due to changes in burst strength and age are much more
significant than those produced by changes in metallicity. This fact is
also observed in the colour-EW(H$\alpha$) diagrams (Figures~\ref{CCD}b
and \ref{CCD}d), especially in the case of sub-solar metallicity
models. It is thus clear that it is in principle possible to infer burst
strengths and ages from these diagrams, but the determination of
metallicities would be very uncertain. 

\begin{table}
\caption[]{Mean colours, H$\alpha$ equivalent widths (\AA), 
and corresponding standard deviations of the mean for the UCM nIR
sample.}
\begin{tabular}{lrccr}
 & n & $\overline{r-J}$\ \ $\sigma$ & $\overline{J-K}$\ \ $\sigma$ &
 $\overline{\mathrm{EW(H}\alpha\mathrm{)}}$\ \ $\sigma$ \\
\hline
\multicolumn{5}{c}{\sc Not corrected for extinction}\\
\hline
Total & 67 & 1.79\ \ 0.05 & 1.06\ \ 0.04 & 80\ \ 7  \\  
\vspace{-0.1cm}
SB-{\it like} & 49 & 1.88\ \ 0.05 & 1.10\ \ 0.05 & 60\ \ 5  \\  
{\tiny (SBN+DANS)} & & & & \\
\vspace{-0.1cm}
HII-{\it like} {\tiny (HIIH+} & 18 & 1.54\ \ 0.09 & 0.96\ \ 0.30 & \ \ 133\ 16  \\  
{\tiny DHIIH+BCD)} & & & & \\
\hline
\multicolumn{5}{c}{\sc Corrected for extinction}\\
\hline
Total & 66 & 1.26\ \ 0.05 & 0.87\ \ 0.06 & 168\ \ 10 \\  
SB-{\it like} & 49 & 1.28\ \ 0.06 & 0.89\ \ 0.09 & 150\ \ 10  \\  
HII-{\it like} & 17 & 1.21\ \ 0.07 & 0.81\ \ 0.35 & 220\ \ 30  \\  
\hline
\label{datamean}
\end{tabular}
\end{table}

\subsection{Determination of the physical properties of the galaxies}
\label{minimization}

For each individual galaxy we have information on its $r-J$ and $J-K$
colours and H$\alpha$ equivalent width. Thus, each galaxy has a point
associated in the $r-J$, $J-K$, 2.5$\times$log\,EW(H$\alpha$)
three-dimensional space. However, due to the calibration and
photometric errors, the uncertainty in these measurements transforms
these points into probability distributions. As in section~\ref{mean}, we
will assume Gaussian probability distributions for the $r-J$, $J-K$
colours and 2.5$\times$log\,EW(H$\alpha$) with the centres and widths
($\sigma$) given in Table~\ref{data}. The evolutionary synthesis
models that we will associate with each galaxy probability
distribution, will also follow different tracks in this
three-dimensional space.

The three-dimensional probability distributions
($r-J$,$J-K$,2.5$\times$log\,EW(H$\alpha$)) have been reproduced using
a Monte Carlo simulation method. A total of 10$^{3}$ data points were
generated in order to reproduce this distribution for each galaxy. No
significant differences were observed using a larger number (e.g., 10$^{4}$)
of input particles. We estimated the model that better reproduces the
colours and EW(H$\alpha$) for each of the 10$^{3}$ test particles
applying a maximum likelihood method. The maximum likelihood estimator
used, ${\cal L}$, includes two colour terms and an EW(H$\alpha$)
term. Thus,
\begin{equation} 
{\cal L}(t,b,Z) = \prod_{n=1}^{3} {1\over \sqrt{2\pi}\Delta C_n}
 \exp\left( - {(c_n-C_n)^2\over 2 \Delta C_n\,^2}\right)
\label{for1}
\end{equation}
where $C_1$, $C_2$ and $C_3$ are the $r-J$ and $J-K$ colours and
2.5$\times$log\,EW(H$\alpha$) and $\Delta C_1$, $\Delta C_2$ and
$\Delta C_3$ are their corresponding errors. The $c_n$ coefficients
are the $r-J$, $J-K$ and 2.5$\times$log\,EW(H$\alpha$) values
predicted by a given model. A similar estimator was employed by
Abraham et al. \shortcite{abraham99} for a sample of intermediate-$z$
HDF \cite{HDF} galaxies.

Finally, we obtained the age, $t$, burst strength, $b$, and
metallicity, $Z$, of the model that maximizes ${\cal L}$ for each test
particle of the ($r-J$,$J-K$,2.5$\times$log\,EW(H$\alpha$))
probability distribution. Therefore, this procedure effectively
provides the ($t$,$b$,$Z$) probability distribution for each input
galaxy.

The resulting ($t$,$b$,$Z$) probability distributions are in many
cases multi-peaked. Instead of analyzing these probability
distributions as a whole, we have studied the clustering pattern
present in the ($t$,$b$,$Z$) solution space. We have used a single
linkage hierarchical clustering method (see Murtagh \& Heck 1987; see
also Appendix), which allows to isolate different solutions in the
($t$,$b$,$Z$) space. We have recovered the three most representative
solution clusters for each galaxy. In Table~\ref{tablafin} we show the
mean properties of those solution clusters with probability higher
than 20~per cent. This probability is computed as the number of test
particles in a given cluster over the total number of test particles
(10$^{3}$). The errors shown in Table~\ref{tablafin} correspond to the
standard deviation of the data for each solution cluster. In those
cases where all the solutions within a cluster yield the same age,
burst strength, metallicity or mass, no errors were given.

The subsequent statistical analysis of each of the ($t$,$b$,$Z$)
clusters indicates that significant correlations between $t$, $b$ and
$Z$ are present. We have performed a principal component analysis
(PCA hereafter; see Morrison 1976; see also Appendix) of the
individual clusters given in Table~\ref{tablafin}. The orientation of
the first PCA component and the contribution of this component to the
total variance within the solution cluster are also given.

\begin{figure}
\psfig{bbllx=33.,bblly=297.,bburx=263.,bbury=766.,file=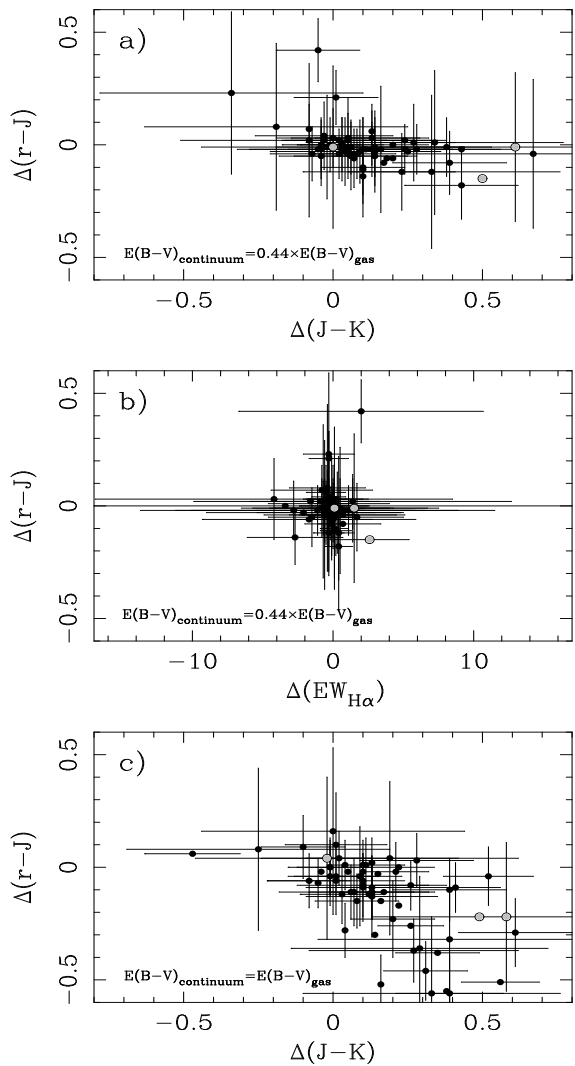,height=14.5cm,width=8cm,angle=0}
\caption{{\bf a)} \& {\bf b)} Differences between the $r-J$, $J-K$
and EW(H$\alpha$) measured values and the best-fitting model
values, using
$E(B-V)_{\mathrm{continuum}}$=0.44$\times$$E(B-V)_{\mathrm{gas}}$. The
input $r-J$, $J-K$ and EW(H$\alpha$) data correspond to the central
values of the respective probability distributions. Those galaxies
with ${\cal L}_{\mathrm{max}}$$<$10 are shown as grey dots. {\bf c)}
Differences in the $r-J$ and $J-K$ colours assuming
$E(B-V)_{\mathrm{continuum}}$=$E(B-V)_{\mathrm{gas}}$.  }
\label{differences}
\end{figure}

After applying this procedure to the observed sample, only three
galaxies (UCM1440$+$2521S, UCM1506$+$1922 and UCM1513$+$2012; see
Figure ~\ref{differences}a and \ref{differences}b) show ${\cal
L}_{\mathrm{max}}$$<$10.0.  Note that a value ${\cal
L}_{\mathrm{max}}=10.0$ corresponds to a model where the differences
between the observed data and the model predictions equal the
measurement errors, assuming mean errors of 0.12$^{\mathrm{m}}$ in
$(r-J)$ and $(J-K)$ and 10~per cent in EW(H$\alpha$). This indicates
that the range of model predictions covers reasonably well the
observed properties of the galaxies. In Figure~\ref{differences}a and
\ref{differences}b we plot the differences between the colours and
EW(H$\alpha$) measured and those predicted by the best-fit model. These
differences have been calculated for the central values of the $r-J$,
$J-K$ and 2.5$\times$log\,EW(H$\alpha$) probability distributions.

Figure~\ref{differences}a shows that, in some cases, the models
predict bluer $J-K$ colours than those measured.  At first sight,
these discrepancies could be explained if we were underestimating the
extinction correction factors applied to these objects. However, since
the extinction correction affects $r-J$ more than $J-K$, applying a
higher extinction correction would destroy the good agreement between
the observed and model $r-J$ colours. We have quantified the effect of
a change in the correction for extinction assumed for our galaxies by
comparing data de-reddened using
$E(B-V)_{\mathrm{continuum}}$=$E(B-V)_{\mathrm{gas}}$ with our models.
Figure~\ref{differences}c shows that data corrected using the relation
given by Calzetti et al. \shortcite{calzetti96} fit the models better
than data corrected assuming
$E(B-V)_{\mathrm{continuum}}$=$E(B-V)_{\mathrm{gas}}$. In addition,
none of the three galaxies given above show higher ${\cal
L}_{\mathrm{max}}$ values after using
$E(B-V)_{\mathrm{continuum}}$=$E(B-V)_{\mathrm{gas}}$. Thus, we are
confident that the extinction correction applied provides a reasonable
fit to the models, and the discrepancies in $J-K$ between the data and
the models are probably due to inherent uncertainties in the modelling
of the nIR continuum by the Bruzual and Charlot code.

The cluster analysis performed indicates that the clustering in the
solution space is basically produced by the discretization in
metallicity of the models. Fortunately, in many of the objects
($\sim$30~per cent; see Table~\ref{tablafin}), only one ($t$,$b$,$Z$)
solution cluster is able to reproduce the observables. About 33~per
cent need two solutions and three solutions are needed for the
remaining 37~per cent. The goodness of this comparison method, given
by the number and size of statistically significant ($t$,$b$,$Z$)
solution clusters, basically depends on the particular position of the
object in the ($r-J$,$J-K$,2.5$\times$log\,EW(H$\alpha$)) space and on
its measurement errors. Fortunately, in those cases where the
($t$,$b$,$Z$) probability distribution is multi-valuated, the
different solution clusters give similar burst strengths and total
stellar masses.  This is another manifestation of the fact that, as we
saw in section~\ref{colcol}, our data is not very sensitive to
metallicity, and we will not attempt to derive it. Nonetheless, the
burst ages are affected somewhat by small changes in metallicity, and
frequently show wider distributions than the burst strenghts.

Finally, the PCA performed on each solution cluster suggests that the
{\it best-axis\/}, given by the vector
($u_t$,$u_{\mathrm{log}(b)}$,$u_{\mathrm{log}(Z)}$)=($u_x$,$u_y$,$u_z$)
shown in Table~\ref{tablafin}, is commonly placed in the $u_x$-$u_y$
($t$-$b$) plane and obeys $u_x$$\simeq$$u_y$. This implies that age
and burst strength are in many cases degenerated and, therefore, the
properties of an individual object can be reproduced both with a
young, low burst strength or an old, high burst strength model, within
the ranges given in Table~\ref{tablafin}.

\begin{table*}
\caption[]{Best fit model results for the nIR sample.}
\begin{tabular}{lcrcrrcc} 
Galaxy & Prob. & Age\ \ \ \ \ \ & log $b$ & log ($Z/Z_{\sun}$) & log ($M/M_{\sun})$\ \ \ \ & PCA & Variance \\
       & (\%)  & (Myr)\ \ \ \  &   ($b$=$M_{\mathrm{young}}$/$M_{\mathrm{total}}$)   &                    &                    &     ($u_t$,$u_{\mathrm{log}\,b}$,$u_{\mathrm{log}\,Z}$) &   (\%)   \\ 
\hline
0003$+$2200  & 39.1 & 6.91$\pm$0.56 & $-$0.981$\pm$0.410 & $-$0.60$\pm$0.14 & 10.18$\pm$0.21(0.14) & ($+$0.666,$+$0.477,$-$0.574) &  73.7 \\
             & 36.7 &11.55$\pm$0.48 & $-$0.813$\pm$0.277 & $-$1.70\,\ \ \ \ \ \ \ \ & 10.44$\pm$0.05(0.14) & ($+$0.707,$+$0.707,$+$0.000) &  62.1 \\
             & 24.2 & 4.55$\pm$0.68 & $-$1.519$\pm$0.380 & $+$0.31$\pm$0.16 & 10.31$\pm$0.27(0.14) & ($+$0.679,$+$0.528,$-$0.510) &  70.0 \\
0013$+$1942  & 43.2 & 4.85$\pm$0.67 & $-$1.846$\pm$0.128 & $-$0.30$\pm$0.32 & 10.10$\pm$0.01(0.03) & ($+$0.634,$+$0.523,$-$0.569) &  82.1 \\
             & 33.5 & 8.86$\pm$0.37 & $-$1.436$\pm$0.103 & $-$1.70\,\ \ \ \ \ \ \ \ & 10.12\,\ \ \ \ \ \ \ (0.03)          & ($+$0.707,$+$0.707,$+$0.000) &  65.7 \\
             & 23.3 & 3.18$\pm$0.10 & $-$1.959$\pm$0.093 & $+$0.40\,\ \ \ \ \ \ \ \ & 10.11$\pm$0.01(0.03) & ($+$0.707,$+$0.707,$+$0.000) &  63.0 \\
0014$+$1748  & 62.8 & 4.33$\pm$0.71 & $-$2.024$\pm$0.192 & $-$0.59$\pm$0.21 & 11.00$\pm$0.01(0.02) & ($+$0.708,$+$0.630,$-$0.319) &  64.7 \\
             & 28.4 & 4.93$\pm$1.43 & $-$2.101$\pm$0.286 & $-$1.70\,\ \ \ \ \ \ \ \ & 11.00$\pm$0.01(0.02) & ($+$0.707,$+$0.707,$+$0.000) &  66.5 \\
0014$+$1829  & 36.8 & 3.31$\pm$0.28 & $-$1.194$\pm$0.188 & $-$0.40$\pm$0.32 & 10.09$\pm$0.08(0.08) & ($+$0.710,$+$0.439,$-$0.550) &  62.6 \\
             & 32.2 & 1.73$\pm$0.30 & $-$1.131$\pm$0.207 & $+$0.40\,\ \ \ \ \ \ \ \          & 10.11$\pm$0.07(0.08) & ($+$0.707,$+$0.707,$+$0.000) &  58.5 \\
             & 31.0 & 3.81$\pm$0.66 & $-$0.885$\pm$0.464 & $-$1.70\,\ \ \ \ \ \ \ \          &  9.87$\pm$0.28(0.08) & ($+$0.707,$+$0.707,$+$0.000) &  56.9 \\
0015$+$2212  & 53.2 & 3.41$\pm$0.81 & $-$2.159$\pm$0.170 & $+$0.23$\pm$0.20 & 10.20$\pm$0.01(0.03) & ($+$0.621,$+$0.561,$-$0.547) &  84.1 \\
             & 24.3 & 4.77$\pm$0.70 & $-$2.135$\pm$0.274 & $-$0.60$\pm$0.14 & 10.19$\pm$0.01(0.03) & ($+$0.623,$+$0.676,$+$0.394) &  71.3 \\
             & 22.5 & 7.71$\pm$1.54 & $-$1.796$\pm$0.318 & $-$1.70\,\ \ \ \ \ \ \ \          & 10.20\,\ \ \ \ \ \ \ (0.03)          & ($+$0.707,$+$0.707,$+$0.000) &  66.5 \\
0017$+$1942  & 67.9 & 4.80$\pm$0.61 & $-$1.613$\pm$0.187 & $-$0.42$\pm$0.31 & 10.51$\pm$0.02(0.03) & ($+$0.719,$+$0.436,$-$0.541) &  63.6 \\
             & 20.6 & 8.32$\pm$0.47 & $-$1.256$\pm$0.144 & $-$1.70\,\ \ \ \ \ \ \ \          & 10.53$\pm$0.01(0.03) & ($+$0.707,$+$0.707,$+$0.000) &  65.3 \\
0022$+$2049  & 42.6 & 1.93$\pm$0.67 & $-$2.214$\pm$0.147 & $+$0.40\,\ \ \ \ \ \ \ \          & 10.98\,\ \ \ \ \ \ \ (0.02)          & ($+$0.707,$+$0.707,$+$0.000) &  63.0 \\
             & 34.6 & 3.32$\pm$1.03 & $-$2.215$\pm$0.249 & $-$0.30$\pm$0.29 & 10.98$\pm$0.01(0.02) & ($+$0.717,$+$0.661,$-$0.224) &  61.8 \\
             & 22.8 & 5.22$\pm$1.80 & $-$2.014$\pm$0.327 & $-$1.70\,\ \ \ \ \ \ \ \          & 10.98$\pm$0.01(0.02) & ($+$0.707,$+$0.707,$+$0.000) &  66.3 \\
0050$+$2114  & 52.2 & 4.76$\pm$0.71 & $-$1.889$\pm$0.224 & $-$0.48$\pm$0.28 & 11.03$\pm$0.01(0.04) & ($+$0.706,$+$0.569,$-$0.421) &  65.9 \\
             & 27.0 & 2.75$\pm$0.37 & $-$2.081$\pm$0.109 & $+$0.40\,\ \ \ \ \ \ \ \          & 11.04$\pm$0.01(0.04) & ($+$0.707,$+$0.707,$+$0.000) &  61.3 \\
             & 20.8 & 7.58$\pm$1.26 & $-$1.643$\pm$0.273 & $-$1.70\,\ \ \ \ \ \ \ \          & 11.04\,\ \ \ \ \ \ \ (0.04)          & ($+$0.707,$+$0.707,$+$0.000) &  66.3 \\
0145$+$2519  & 72.1 &11.56$\pm$0.17 & $-$0.979$\pm$0.107 & $-$1.70\,\ \ \ \ \ \ \ \          & 11.30$\pm$0.02(0.04) & ($+$0.707,$+$0.707,$+$0.000) &  64.5 \\
1255$+$3125  & 63.7 & 4.78$\pm$0.99 & $-$1.423$\pm$0.650 & $+$0.34$\pm$0.14 & 10.28$\pm$0.45(0.07) & ($+$0.618,$+$0.683,$+$0.389) &  67.9 \\
             & 22.3 & 5.13$\pm$1.36 & $-$2.144$\pm$0.397 & $-$0.55$\pm$0.15 & 10.71$\pm$0.04(0.07) & ($+$0.688,$+$0.702,$+$0.182) &  65.2 \\
1256$+$2823  & 66.5 & 3.65$\pm$0.45 & $-$1.701$\pm$0.100 & $+$0.35$\pm$0.13 & 10.88$\pm$0.01(0.04) & ($+$0.683,$+$0.369,$-$0.631) &  70.8 \\
             & 28.0 & 9.84$\pm$0.43 & $-$1.162$\pm$0.140 & $-$1.70\,\ \ \ \ \ \ \ \          & 10.89$\pm$0.01(0.04) & ($+$0.707,$+$0.707,$+$0.000) &  66.3 \\
1257$+$2808  & 59.6 &11.46$\pm$0.18 & $-$0.405$\pm$0.148 & $-$1.70\,\ \ \ \ \ \ \ \          & 10.22$\pm$0.03(0.11) & ($+$0.707,$+$0.707,$+$0.000) &  64.2 \\
             & 38.5 & 4.47$\pm$0.54 & $-$1.138$\pm$0.249 & $+$0.32$\pm$0.16 & 10.10$\pm$0.19(0.11) & ($+$0.650,$+$0.484,$-$0.586) &  77.9 \\
1259$+$2755  & 85.1 & 4.35$\pm$0.50 & $-$1.234$\pm$0.240 & $+$0.34$\pm$0.14 & 10.71$\pm$0.17(0.05) & ($+$0.658,$+$0.488,$-$0.574) &  76.4 \\
1259$+$3011  & 84.1 &13.84$\pm$0.32 & $-$0.923$\pm$0.146 & $-$1.70\,\ \ \ \ \ \ \ \          & 10.79$\pm$0.03(0.06) & ($+$0.707,$+$0.707,$+$0.000) &  65.7 \\
1302$+$2853  & 57.7 & 6.98$\pm$0.55 & $-$1.471$\pm$0.271 & $-$0.60$\pm$0.14 & 10.40$\pm$0.12(0.08) & ($+$0.700,$+$0.380,$-$0.605) &  66.4 \\
             & 25.4 &11.65$\pm$0.36 & $-$1.154$\pm$0.148 & $-$1.70\,\ \ \ \ \ \ \ \         & 10.51$\pm$0.01(0.08) & ($+$0.707,$+$0.707,$+$0.000) &  65.7 \\
1304$+$2808  & 82.6 &13.96$\pm$0.96 & $-$1.487$\pm$0.216 & $-$1.70\,\ \ \ \ \ \ \ \          & 10.71$\pm$0.01(0.06) & ($+$0.707,$+$0.707,$+$0.000) &  66.3 \\
1304$+$2818  & 89.0 & 3.47$\pm$0.28 & $-$1.988$\pm$0.143 & $+$0.39$\pm$0.06 & 10.68$\pm$0.01(0.04) & ($+$0.720,$+$0.656,$-$0.225) &  61.1 \\
1306$+$2938  & 66.8 & 3.37$\pm$0.22 & $-$1.746$\pm$0.096 & $+$0.39$\pm$0.07 & 10.67$\pm$0.01(0.04) & ($+$0.711,$+$0.309,$-$0.632) &  65.4 \\
             & 32.0 & 9.31$\pm$0.27 & $-$1.248$\pm$0.087 & $-$1.70\,\ \ \ \ \ \ \ \          & 10.68\,\ \ \ \ \ \ \ (0.04)          & ($+$0.707,$+$0.707,$+$0.000) &  65.0 \\
1307$+$2910  & 46.4 &12.38$\pm$0.77 & $-$0.884$\pm$0.317 & $-$1.70\,\ \ \ \ \ \ \ \         & 11.25$\pm$0.07(0.11) & ($+$0.707,$+$0.707,$+$0.000) &  61.5 \\
             & 46.3 & 6.97$\pm$1.07 & $-$1.157$\pm$0.537 & $-$0.44$\pm$0.29 & 11.02$\pm$0.26(0.11) & ($+$0.690,$+$0.387,$-$0.612) &  66.7 \\
1308$+$2950  & 39.7 & 5.21$\pm$0.59 & $-$1.559$\pm$0.131 & $-$0.38$\pm$0.30 & 11.40$\pm$0.02(0.04) & ($+$0.631,$+$0.505,$-$0.589) &  82.0 \\
             & 35.2 & 9.19$\pm$0.27 & $-$1.170$\pm$0.094 & $-$1.70\,\ \ \ \ \ \ \ \          & 11.43\,\ \ \ \ \ \ \ (0.04)          & ($+$0.707,$+$0.707,$+$0.000) &  65.9 \\
             & 25.1 & 3.25$\pm$0.08 & $-$1.695$\pm$0.099 & $+$0.40\,\ \ \ \ \ \ \ \          & 11.41$\pm$0.01(0.04) & ($+$0.707,$+$0.707,$+$0.000) &  65.6 \\
1308$+$2958  & 37.0 & 7.60$\pm$0.06 & $-$0.738$\pm$0.146 & $-$0.70\,\ \ \ \ \ \ \ \          & 10.47$\pm$0.11(0.06) & ($+$0.707,$+$0.707,$+$0.000) &  61.6 \\
             & 35.7 & 5.75\,\ \ \ \ \ \ \ \ & $-$0.717$\pm$0.101 & $+$0.00\,\ \ \ \ \ \ \ \          & 10.38$\pm$0.07(0.06) & ($+$0.000,$+$1.000,$+$0.000) & 100.0 \\
             & 27.3 &12.00\,\ \ \ \ \ \ \ \ & $-$0.647$\pm$0.038 & $-$1.70\,\ \ \ \ \ \ \ \          & 10.76$\pm$0.01(0.06) & ($+$0.000,$+$1.000,$+$0.000) & 100.0 \\
1312$+$2954  & 39.8 & 5.89$\pm$0.28 & $-$1.438$\pm$0.208 & $-$0.54$\pm$0.15 & 10.71$\pm$0.09(0.13) & ($+$0.743,$+$0.207,$-$0.636) &  56.5 \\
             & 39.2 & 3.79$\pm$0.55 & $-$1.642$\pm$0.172 & $+$0.33$\pm$0.15 & 10.76$\pm$0.05(0.13) & ($+$0.675,$+$0.408,$-$0.615) &  71.9 \\
             & 21.0 & 9.99$\pm$0.56 & $-$1.114$\pm$0.171 & $-$1.70\,\ \ \ \ \ \ \ \         & 10.79$\pm$0.01(0.13) & ($+$0.707,$+$0.707,$+$0.000) &  64.8 \\
1312$+$3040  & 73.4 & 3.69$\pm$0.39 & $-$2.073$\pm$0.181 & $+$0.37$\pm$0.10 & 10.86$\pm$0.05(0.03) & ($+$0.739,$+$0.540,$-$0.403) &  60.5 \\
             & 25.2 & 9.87$\pm$0.63 & $-$1.577$\pm$0.144 & $-$1.70\,\ \ \ \ \ \ \ \          & 10.88\,\ \ \ \ \ \ \ (0.03)          & ($+$0.707,$+$0.707,$+$0.000) &  66.4 \\
1428$+$2727  & 53.1 & 3.49$\pm$0.13 & $-$1.400$\pm$0.145 & $+$0.40\,\ \ \ \ \ \ \ \          & 10.17$\pm$0.03(0.07) & ($+$0.707,$+$0.707,$+$0.000) &  65.4 \\
             & 25.6 & 9.97$\pm$0.41 & $-$0.809$\pm$0.209 & $-$1.70\,\ \ \ \ \ \ \ \          & 10.20$\pm$0.03(0.07) & ($+$0.707,$+$0.707,$+$0.000) &  66.1 \\
             & 21.3 & 5.53$\pm$0.49 & $-$1.146$\pm$0.268 & $-$0.37$\pm$0.26 & 10.07$\pm$0.12(0.07) & ($+$0.663,$+$0.447,$-$0.600) &  74.6 \\
1432$+$2645  & 65.6 &11.49$\pm$0.07 & $-$0.974$\pm$0.042 & $-$1.70\,\ \ \ \ \ \ \ \          & 11.14\,\ \ \ \ \ \ \ (0.04)          & ($+$0.707,$+$0.707,$+$0.000) &  60.8 \\
             & 25.9 & 4.50$\pm$0.52 & $-$1.610$\pm$0.079 & $+$0.32$\pm$0.16 & 11.08$\pm$0.03(0.04) & ($+$0.617,$+$0.527,$-$0.585) &  86.0 \\
1440$+$2511  & 96.7 &12.60$\pm$0.03 & $-$0.882$\pm$0.030 & $-$1.70\,\ \ \ \ \ \ \ \          & 10.78\,\ \ \ \ \ \ \ (0.05)          & ($+$0.707,$+$0.707,$-$0.004) &  47.9 \\
1440$+$2521N & 42.8 & 3.58$\pm$0.74 & $-$1.724$\pm$0.502 & $+$0.40\,\ \ \ \ \ \ \ \          & 10.80$\pm$0.25(0.12) & ($+$0.707,$+$0.707,$+$0.000) &  64.6 \\
             & 33.2 & 4.95$\pm$1.20 & $-$1.871$\pm$0.458 & $-$0.33$\pm$0.29 & 10.86$\pm$0.08(0.12) & ($+$0.704,$+$0.664,$-$0.250) &  64.6 \\
             & 24.0 & 8.09$\pm$3.17 & $-$1.660$\pm$0.661 & $-$1.70\,\ \ \ \ \ \ \ \          & 10.90$\pm$0.03(0.12) & ($+$0.707,$+$0.707,$+$0.000) &  65.8 \\
1440$+$2521S & 61.6 & 3.88$\pm$0.66 & $-$1.788$\pm$0.352 & $+$0.36$\pm$0.11 & 10.52$\pm$0.15(0.12) & ($+$0.720,$+$0.671,$-$0.178) &  62.2 \\
             & 31.8 & 9.89$\pm$2.78 & $-$1.295$\pm$0.621 & $-$1.70\,\ \ \ \ \ \ \ \          & 10.57$\pm$0.03(0.12) & ($+$0.707,$+$0.707,$+$0.000) &  65.6 \\
1442$+$2845  & 59.0 & 4.55$\pm$0.63 & $-$1.986$\pm$0.193 & $-$0.41$\pm$0.31 & 10.32$\pm$0.01(0.04) & ($+$0.738,$+$0.532,$-$0.415) &  60.6 \\
             & 20.6 & 7.47$\pm$1.42 & $-$1.705$\pm$0.302 & $-$1.70\,\ \ \ \ \ \ \ \          & 10.32\,\ \ \ \ \ \ \ (0.04)          & ($+$0.707,$+$0.707,$+$0.000) &  66.4 \\
             & 20.4 & 2.70$\pm$0.42 & $-$2.149$\pm$0.130 & $+$0.40\,\ \ \ \ \ \ \ \          & 10.32$\pm$0.01(0.04) & ($+$0.707,$+$0.707,$+$0.000) &  63.2 \\
\hline
\end{tabular}
\end{table*}

\begin{table*}
\addtocounter{table}{-1}
\caption[]{(cont.) Best fit model results for the nIR sample.}
\begin{tabular}{lcrcrrcc} 
Galaxy & Prob. & Age\ \ \ \ \ \ & log $b$ & log ($Z/Z_{\sun}$) &  log ($M/M_{\sun}$)\ \ \ \ & PCA & Variance \\
       & (\%)  & (Myr)\ \ \ \  &  ($b$=$M_{\mathrm{young}}$/$M_{\mathrm{total}}$)   &                    &                    &  ($u_t$,$u_{\mathrm{log}\,b}$,$u_{\mathrm{log}\,Z}$)   &   (\%)   \\ 
\hline 
1443$+$2548  & 62.3 & 5.00$\pm$1.47 & $-$1.767$\pm$0.572 & $-$0.40$\pm$0.26 & 10.89$\pm$0.10(0.10) & ($+$0.702,$+$0.710,$+$0.061) &  62.4 \\
             & 24.0 & 4.89$\pm$3.56 & $-$2.275$\pm$0.706 & $-$1.70\,\ \ \ \ \ \ \ \          & 10.98$\pm$0.03(0.10) & ($+$0.707,$+$0.707,$+$0.000) &  65.7 \\
1452$+$2754  & 52.3 & 3.13$\pm$1.39 & $-$2.398$\pm$0.462 & $-$0.57$\pm$0.24 & 11.13$\pm$0.04(0.10) & ($+$0.673,$+$0.704,$+$0.226) &  65.8 \\
             & 37.8 & 2.52$\pm$2.14 & $-$2.573$\pm$0.367 & $-$1.70\,\ \ \ \ \ \ \ \          & 11.14$\pm$0.01(0.10) & ($+$0.707,$+$0.707,$+$0.000) &  65.5 \\
1506$+$1922  & 46.6 & 2.81$\pm$2.98 & $-$2.690$\pm$0.518 & $-$1.70\,\ \ \ \ \ \ \ \          & 10.82$\pm$0.01(0.10) & ($+$0.707,$+$0.707,$+$0.000) &  66.1 \\
             & 33.1 & 4.05$\pm$1.68 & $-$2.252$\pm$0.559 & $-$0.44$\pm$0.29 & 10.78$\pm$0.10(0.10) & ($+$0.698,$+$0.708,$+$0.107) &  63.4 \\
             & 20.3 & 3.55$\pm$1.36 & $-$1.675$\pm$0.894 & $+$0.40\,\ \ \ \ \ \ \ \          & 10.46$\pm$0.56(0.10) & ($+$0.707,$+$0.707,$+$0.000) &  64.6 \\
1513$+$2012  & 90.3 & 2.75$\pm$0.59 & $-$2.137$\pm$0.117 & $+$0.31$\pm$0.17 & 11.29\,\ \ \ \ \ \ \ (0.03)          & ($+$0.644,$+$0.508,$-$0.572) &  76.5 \\
1557$+$1423  & 42.5 &11.71$\pm$0.41 & $-$1.450$\pm$0.118 & $-$1.70\,\ \ \ \ \ \ \ \          & 10.61$\pm$0.01(0.02) & ($+$0.707,$+$0.707,$+$0.000) &  66.2 \\
             & 34.6 & 4.39$\pm$0.56 & $-$2.167$\pm$0.162 & $+$0.32$\pm$0.16 & 10.59$\pm$0.03(0.02) & ($+$0.695,$+$0.499,$-$0.518) &  67.9 \\
             & 22.9 & 7.16$\pm$0.54 & $-$1.824$\pm$0.165 & $-$0.64$\pm$0.12 & 10.57$\pm$0.03(0.02) & ($+$0.651,$+$0.530,$-$0.544) &  77.8 \\
1646$+$2725  & 35.2 & 2.89$\pm$1.43 & $-$2.231$\pm$0.329 & $-$0.60$\pm$0.14 &  9.93$\pm$0.02(0.05) & ($+$0.688,$+$0.693,$+$0.214) &  64.3 \\
             & 34.2 & 3.00$\pm$2.20 & $-$2.248$\pm$0.365 & $-$1.70\,\ \ \ \ \ \ \ \          &  9.93$\pm$0.01(0.05) & ($+$0.707,$+$0.707,$+$0.000) &  66.3 \\
             & 30.6 & 1.78$\pm$0.91 & $-$2.179$\pm$0.187 & $+$0.28$\pm$0.18 &  9.94$\pm$0.01(0.05) & ($+$0.720,$+$0.681,$-$0.133) &  59.1 \\
1647$+$2729  & 66.4 & 4.66$\pm$0.59 & $-$1.090$\pm$0.668 & $+$0.31$\pm$0.17 & 10.53$\pm$0.51(0.02) & ($+$0.514,$+$0.764,$+$0.390) &  56.6 \\
1647$+$2950  & 38.2 & 4.87$\pm$1.29 & $-$1.769$\pm$0.515 & $-$0.57$\pm$0.15 & 11.02$\pm$0.08(0.11) & ($+$0.635,$+$0.679,$+$0.369) &  68.1 \\
             & 36.6 & 3.45$\pm$1.05 & $-$1.881$\pm$0.385 & $+$0.19$\pm$0.20 & 11.05$\pm$0.06(0.11) & ($+$0.688,$+$0.584,$-$0.431) &  66.5 \\
             & 25.2 & 6.21$\pm$3.25 & $-$1.848$\pm$0.663 & $-$1.70\,\ \ \ \ \ \ \ \          & 11.08$\pm$0.01(0.11) & ($+$0.707,$+$0.707,$+$0.000) &  66.2 \\
1648$+$2855  & 92.2 & 2.71$\pm$0.47 & $-$1.777$\pm$0.085 & $+$0.35$\pm$0.14 & 10.78$\pm$0.01(0.05) & ($+$0.623,$+$0.518,$-$0.586) &  78.7 \\
1654$+$2812  & 60.9 & 4.30$\pm$0.62 & $-$1.756$\pm$0.242 & $+$0.34$\pm$0.14 &  9.92$\pm$0.14(0.07) & ($+$0.697,$+$0.528,$-$0.486) &  67.8 \\
             & 20.2 & 6.43$\pm$0.47 & $-$1.577$\pm$0.211 & $-$0.53$\pm$0.15 &  9.91$\pm$0.07(0.07) & ($+$0.692,$+$0.447,$-$0.567) &  68.1 \\
1656$+$2744  & 55.6 & 4.45$\pm$1.13 & $-$2.164$\pm$0.295 & $-$0.30$\pm$0.26 & 10.72$\pm$0.01(0.04) & ($+$0.666,$+$0.628,$-$0.403) &  71.4 \\
             & 22.2 & 8.15$\pm$1.33 & $-$1.773$\pm$0.283 & $-$1.70\,\ \ \ \ \ \ \ \          & 10.72\,\ \ \ \ \ \ \ (0.04)          & ($+$0.707,$+$0.707,$+$0.000) &  66.3 \\
             & 22.2 & 2.82$\pm$0.94 & $-$2.174$\pm$0.389 & $+$0.40\,\ \ \ \ \ \ \ \          & 10.70$\pm$0.14(0.04) & ($+$0.707,$+$0.707,$+$0.000) &  63.4 \\
1657$+$2901  & 84.3 & 3.94$\pm$0.40 & $-$1.760$\pm$0.190 & $+$0.38$\pm$0.09 & 10.52$\pm$0.07(0.05) & ($+$0.703,$+$0.526,$-$0.479) &  67.3 \\
2238$+$2308  & 63.8 & 5.67$\pm$0.21 & $-$1.593$\pm$0.106 & $-$0.65$\pm$0.11 & 11.22$\pm$0.02(0.02) & ($+$0.749,$+$0.169,$-$0.640) &  54.6 \\
             & 23.7 & 3.89$\pm$0.73 & $-$1.664$\pm$0.142 & $+$0.23$\pm$0.20 & 11.22$\pm$0.01(0.02) & ($+$0.615,$+$0.527,$-$0.586) &  86.2 \\
2239$+$1959  & 68.8 & 3.19$\pm$0.65 & $-$1.895$\pm$0.130 & $+$0.25$\pm$0.19 & 11.11$\pm$0.01(0.02) & ($+$0.606,$+$0.556,$-$0.569) &  88.3 \\
2250$+$2427  & 87.4 & 2.27$\pm$0.66 & $-$2.056$\pm$0.131 & $+$0.32$\pm$0.16 & 11.49$\pm$0.01(0.02) & ($+$0.714,$+$0.466,$-$0.523) &  63.2 \\
2251$+$2352  & 51.5 & 4.27$\pm$0.40 & $-$1.943$\pm$0.113 & $+$0.38$\pm$0.09 & 10.39$\pm$0.04(0.02) & ($+$0.683,$+$0.533,$-$0.499) &  70.9 \\
             & 35.4 &11.40$\pm$0.48 & $-$1.304$\pm$0.139 & $-$1.70\,\ \ \ \ \ \ \ \          & 10.43$\pm$0.01(0.02) & ($+$0.707,$+$0.707,$+$0.000) &  66.2 \\
2253$+$2219  & 48.0 & 4.65$\pm$0.69 & $-$2.152$\pm$0.149 & $-$0.21$\pm$0.30 & 10.72$\pm$0.01(0.02) & ($+$0.643,$+$0.536,$-$0.546) &  80.3 \\
             & 31.3 & 8.49$\pm$0.73 & $-$1.794$\pm$0.167 & $-$1.70\,\ \ \ \ \ \ \ \          & 10.72\,\ \ \ \ \ \ \ (0.02)          & ($+$0.707,$+$0.707,$+$0.000) &  66.6 \\
             & 20.7 & 3.46$\pm$0.95 & $-$1.945$\pm$0.757 & $+$0.40\,\ \ \ \ \ \ \ \          & 10.50$\pm$0.51(0.02) & ($+$0.707,$+$0.707,$+$0.000) &  65.6 \\
2255$+$1654  & 63.0 & 3.79$\pm$0.67 & $-$1.963$\pm$0.116 & $+$0.21$\pm$0.20 & 11.51$\pm$0.01(0.02) & ($+$0.649,$+$0.475,$-$0.594) &  77.6 \\
             & 30.3 & 9.26$\pm$0.22 & $-$1.401$\pm$0.062 & $-$1.70\,\ \ \ \ \ \ \ \          & 11.52\,\ \ \ \ \ \ \ (0.02)          & ($+$0.707,$+$0.707,$+$0.000) &  66.6 \\
2255$+$1926  & 73.2 &13.08$\pm$0.34 & $-$1.135$\pm$0.133 & $-$1.70\,\ \ \ \ \ \ \ \          & 10.02$\pm$0.01(0.03) & ($+$0.707,$+$0.707,$+$0.000) &  65.9 \\
2255$+$1930N & 65.5 & 3.53$\pm$0.74 & $-$2.102$\pm$0.152 & $+$0.18$\pm$0.20 & 10.86$\pm$0.01(0.02) & ($+$0.643,$+$0.530,$-$0.552) &  79.6 \\
             & 20.1 & 5.02$\pm$0.55 & $-$2.014$\pm$0.225 & $-$0.54$\pm$0.15 & 10.85$\pm$0.01(0.02) & ($+$0.679,$+$0.709,$+$0.192) &  65.5 \\
2255$+$1930S & 49.9 & 4.15$\pm$0.39 & $-$1.955$\pm$0.149 & $+$0.40\,\ \ \ \ \ \ \ \          & 10.38$\pm$0.06(0.02) & ($+$0.707,$+$0.707,$+$0.000) &  66.0 \\
             & 28.7 & 6.15$\pm$0.73 & $-$1.853$\pm$0.196 & $-$0.42$\pm$0.30 & 10.39$\pm$0.03(0.02) & ($+$0.661,$+$0.500,$-$0.559) &  75.8 \\
             & 21.4 &10.97$\pm$0.57 & $-$1.428$\pm$0.154 & $-$1.70\,\ \ \ \ \ \ \ \          & 10.42$\pm$0.01(0.02) & ($+$0.707,$+$0.707,$+$0.000) &  66.4 \\
2258$+$1920  & 67.8 & 1.32$\pm$0.92 & $-$2.692$\pm$0.102 & $-$1.70\,\ \ \ \ \ \ \ \          & 10.65\,\ \ \ \ \ \ \ (0.02)          & ($+$0.707,$+$0.707,$+$0.000) &  61.1 \\
             & 24.0 & 1.52$\pm$1.12 & $-$2.576$\pm$0.115 & $-$0.36$\pm$0.25 & 10.65\,\ \ \ \ \ \ \ (0.02)          & ($+$0.643,$+$0.708,$+$0.293) &  60.6 \\
2300$+$2015  & 77.3 & 1.27$\pm$0.58 & $-$2.713$\pm$0.062 & $-$1.70\,\ \ \ \ \ \ \ \          & 10.90\,\ \ \ \ \ \ \ (0.02)          & ($+$0.707,$+$0.707,$+$0.000) &  56.5 \\
2302$+$2053E & 79.5 & 4.34$\pm$0.47 & $-$1.349$\pm$0.123 & $+$0.35$\pm$0.14 & 11.23$\pm$0.06(0.02) & ($+$0.613,$+$0.532,$-$0.584) &  86.7 \\
2302$+$2053W & 49.4 & 2.38$\pm$0.95 & $-$2.239$\pm$0.142 & $-$0.38$\pm$0.27 & 10.23$\pm$0.01(0.03) & ($+$0.734,$+$0.605,$-$0.308) &  57.8 \\
             & 26.3 & 1.26$\pm$0.63 & $-$2.117$\pm$0.096 & $+$0.40\,\ \ \ \ \ \ \ \          & 10.23\,\ \ \ \ \ \ \ (0.03)          & ($+$0.707,$+$0.707,$+$0.000) &  58.3 \\
             & 24.3 & 3.65$\pm$1.20 & $-$2.077$\pm$0.253 & $-$1.70\,\ \ \ \ \ \ \ \          & 10.21$\pm$0.01(0.03) & ($+$0.707,$+$0.707,$+$0.000) &  65.2 \\
2303$+$1856  & 55.5 & 3.47$\pm$0.72 & $-$1.856$\pm$0.164 & $+$0.19$\pm$0.20 & 11.27$\pm$0.01(0.02) & ($+$0.608,$+$0.554,$-$0.569) &  88.7 \\
             & 23.1 & 8.40$\pm$0.55 & $-$1.334$\pm$0.155 & $-$1.70\,\ \ \ \ \ \ \ \          & 11.28\,\ \ \ \ \ \ \ (0.02)          & ($+$0.707,$+$0.707,$+$0.000) &  66.1 \\
             & 21.4 & 5.09$\pm$0.40 & $-$1.721$\pm$0.170 & $-$0.55$\pm$0.15 & 11.26$\pm$0.01(0.02) & ($+$0.710,$+$0.691,$-$0.136) &  64.9 \\
2304$+$1640  & 47.6 & 4.97$\pm$0.65 & $-$1.587$\pm$0.186 & $-$0.33$\pm$0.31 &  9.43$\pm$0.03(0.06) & ($+$0.647,$+$0.502,$-$0.574) &  78.5 \\
             & 31.7 & 8.90$\pm$0.51 & $-$1.198$\pm$0.175 & $-$1.70\,\ \ \ \ \ \ \ \          &  9.45$\pm$0.01(0.06) & ($+$0.707,$+$0.707,$+$0.000) &  66.1 \\
             & 20.7 & 3.17$\pm$0.17 & $-$1.710$\pm$0.153 & $+$0.40\,\ \ \ \ \ \ \ \          &  9.44$\pm$0.02(0.06) & ($+$0.707,$+$0.707,$+$0.000) &  62.3 \\
2307$+$1947  & 50.0 & 4.21$\pm$1.50 & $-$2.355$\pm$0.516 & $+$0.28$\pm$0.18 & 10.73$\pm$0.18(0.03) & ($+$0.680,$+$0.704,$+$0.204) &  65.7 \\
             & 33.2 & 8.59$\pm$4.48 & $-$2.173$\pm$0.802 & $-$1.70\,\ \ \ \ \ \ \ \          & 10.81$\pm$0.01(0.03) & ($+$0.707,$+$0.707,$+$0.000) &  66.5 \\
2310$+$1800  & 45.2 & 4.46$\pm$1.14 & $-$2.230$\pm$0.271 & $-$0.34$\pm$0.29 & 11.15$\pm$0.01(0.03) & ($+$0.714,$+$0.673,$-$0.193) &  63.2 \\
             & 28.2 & 3.10$\pm$1.22 & $-$2.072$\pm$0.678 & $+$0.40\,\ \ \ \ \ \ \ \          & 10.99$\pm$0.36(0.03) & ($+$0.707,$+$0.707,$+$0.000) &  64.4 \\
             & 26.6 & 6.98$\pm$2.85 & $-$2.032$\pm$0.520 & $-$1.70\,\ \ \ \ \ \ \ \          & 11.15\,\ \ \ \ \ \ \ (0.03)          & ($+$0.707,$+$0.707,$+$0.000) &  66.4 \\
2313$+$1841  & 70.2 & 5.03$\pm$0.65 & $-$1.777$\pm$0.184 & $-$0.43$\pm$0.29 & 10.65$\pm$0.02(0.04) & ($+$0.675,$+$0.510,$-$0.534) &  72.5 \\
2316$+$2028  & 99.8 & 0.92$\pm$0.36 & $-$2.731$\pm$0.026 & $-$1.70\,\ \ \ \ \ \ \ \          & 10.65\,\ \ \ \ \ \ \ (0.04)          & ($+$0.707,$+$0.707,$+$0.002) &  39.0 \\
2316$+$2457  & 43.0 & 4.69$\pm$0.65 & $-$2.069$\pm$0.233 & $-$0.62$\pm$0.13 & 11.63$\pm$0.01(0.04) & ($+$0.703,$+$0.709,$+$0.059) &  65.1 \\
\hline
\end{tabular}
\end{table*}

\begin{table*}
\addtocounter{table}{-1}
\caption[]{(cont.) Best fit model results for the nIR sample.}
\begin{tabular}{lcrcrrcc} 
Galaxy & Prob. & Age\ \ \ \ \ \ & log $b$ & log ($Z/Z_{\sun}$) &  log ($M/M_{\sun}$)\ \ \ \ & PCA & Variance \\
       & (\%)  & (Myr)\ \ \ \  &  ($b$=$M_{\mathrm{young}}$/$M_{\mathrm{total}}$)   &                    &                    &  ($u_t$,$u_{\mathrm{log}\,b}$,$u_{\mathrm{log}\,Z}$)   &   (\%)   \\ 
\hline 
2316$+$2457  & 35.5 & 2.96$\pm$0.89 & $-$2.167$\pm$0.232 & $+$0.24$\pm$0.20 & 11.63$\pm$0.01(0.04) & ($+$0.679,$+$0.599,$-$0.424) &  68.9 \\
             & 21.5 & 6.60$\pm$1.81 & $-$1.907$\pm$0.343 & $-$1.70\,\ \ \ \ \ \ \ \           & 11.64\,\ \ \ \ \ \ \ (0.04)          & ($+$0.707,$+$0.707,$+$0.000) &  66.4 \\
2316$+$2459  & 71.6 & 4.90$\pm$1.28 & $-$1.256$\pm$0.794 & $+$0.36$\pm$0.12 & 10.37$\pm$0.50(0.04) & ($+$0.616,$+$0.652,$+$0.443) &  73.4 \\
2319$+$2234  & 81.1 & 1.37$\pm$0.37 & $-$2.838$\pm$0.030 & $-$1.70\,\ \ \ \ \ \ \ \           & 10.85\,\ \ \ \ \ \ \ (0.04)          & ($+$0.707,$+$0.707,$+$0.000) &  52.0 \\
2321$+$2149  & 61.0 & 4.51$\pm$0.57 & $-$1.617$\pm$0.219 & $+$0.33$\pm$0.15 & 10.62$\pm$0.15(0.04) & ($+$0.715,$+$0.488,$-$0.500) &  64.7 \\
             & 30.6 & 6.40$\pm$0.36 & $-$1.553$\pm$0.226 & $-$0.56$\pm$0.15 & 10.65$\pm$0.09(0.04) & ($+$0.733,$+$0.152,$-$0.663) &  57.7 \\
2324$+$2448  & 57.3 & 6.96$\pm$2.30 & $-$1.717$\pm$1.105 & $+$0.33$\pm$0.16 & 10.77$\pm$0.43(0.04) & ($+$0.648,$+$0.692,$+$0.317) &  67.3 \\
             & 25.0 &13.51$\pm$2.61 & $-$2.229$\pm$0.446 & $-$1.70\,\ \ \ \ \ \ \ \           & 11.24\,\ \ \ \ \ \ \ (0.04)          & ($+$0.707,$+$0.707,$+$0.000) &  66.3 \\
2327$+$2515N & 35.7 & 5.89$\pm$0.99 & $-$1.391$\pm$0.209 & $-$1.70\,\ \ \ \ \ \ \ \           & 10.25$\pm$0.01(0.07) & ($+$0.707,$+$0.707,$+$0.000) &  66.1 \\
             & 33.1 & 2.28$\pm$0.30 & $-$1.602$\pm$0.125 & $+$0.40\,\ \ \ \ \ \ \ \           & 10.24$\pm$0.01(0.07) & ($+$0.707,$+$0.707,$+$0.000) &  60.6 \\
             & 31.2 & 3.94$\pm$0.43 & $-$1.548$\pm$0.179 & $-$0.42$\pm$0.29 & 10.22$\pm$0.02(0.07) & ($+$0.741,$+$0.473,$-$0.476) &  58.6 \\
2327$+$2515S & 81.9 & 4.25$\pm$0.46 & $-$1.253$\pm$0.370 &    0.37$\pm$0.11 & 10.15$\pm$0.29(0.06) & ($+$0.711,$+$0.494,$-$0.501) &  65.5 \\
\hline
\label{tablafin}
\end{tabular}
\end{table*}

\begin{table}
\caption[]{Mean properties and standard deviations 
for the nIR sample. {\it Dwarfs} includes the DHIIH and BCD spectroscopic type galaxies.}
\begin{tabular}{lrcccc}
 & n & $\overline{t}$\ \ \ $\sigma$ & $\!\!\!\overline{\mathrm{log}b}$\ \ \ \ $\sigma$ & $\!\!\!\overline{\mathrm{log}Z}$\ \ $\sigma$ &
 $\!\!\!\!\!\!\overline{\mathrm{log}M}$\ \ \ \ $\!\sigma$ \\
\hline 
     &  & (Myr) & & &  \\
\hline
\multicolumn{6}{c}{\sc with $E(B-V)_{\mathrm{continuum}}$=0.44$\times$$E(B-V)_{\mathrm{gas}}$}\\
\hline
Total & 67 & $\!$5.5~\ 0.4 & $\!\!\!\!\!$$-$1.72~\  0.07 & $\!\!\!$$-$0.5~\  0.1 & $\!\!$10.69~\  0.06 \\ 
SBN   & 41 & $\!$5.8~\  0.5 & $\!\!\!\!\!$$-$1.69~\  0.10 & $\!\!\!$$-$0.5~\  0.1 & $\!\!$10.90~\  0.06 \\ 
DANS  &  8 & $\!$5.7~\  1.6 & $\!\!\!\!\!$$-$1.91~\  0.23 & $\!\!\!$$-$0.8~\  0.3 & $\!\!$10.57~\  0.06 \\ 
HIIH  & 13 & $\!$4.1~\  0.6 & $\!\!\!\!\!$$-$1.72~\  0.15 & $\!\!\!$$-$0.3~\  0.2 & $\!\!$10.42~\  0.10 \\ 
{\it Dwarfs} &  5 & $\!$6.6~\  1.7 & $\!\!\!\!\!$$-$1.62~\  0.20 & $\!\!\!$$-$0.7~\  0.4 & \ \ $\!\!\!$9.93~\  0.15 \\
\hline
SB    & 49 & $\!$5.8~\  0.5 & $\!\!\!\!\!$$-$1.73~\  0.09 & $\!\!\!$$-$0.6~\  0.1 & $\!\!$10.85~\  0.05 \\  
HII   & 18 & $\!$4.7~\  0.7 & $\!\!\!\!\!$$-$1.69~\  0.10 & $\!\!\!$$-$0.4~\  0.2 & $\!\!$10.29~\  0.10  \\ 
\hline
\multicolumn{6}{c}{\sc with $E(B-V)_{\mathrm{continuum}}=E(B-V)_{\mathrm{gas}}$}\\
\hline
Total & 67 & $\!\!\!$11.5~\ 0.6 & $\!\!\!\!\!$$-$0.77~\  0.07 & $\!\!\!$$-$1.2~\  0.1 & $\!\!$10.64~\  0.05 \\ 
SB    & 49 & $\!\!\!$12.5~\ 0.6 & $\!\!\!\!\!$$-$0.69~\  0.08 & $\!\!\!$$-$1.4~\  0.1 & $\!\!$10.77~\  0.05 \\  
HII   & 18 & $\!$8.3~\ 1.0 & $\!\!\!\!\!$$-$1.01~\  0.11 & $\!\!\!$$-$0.7~\  0.2 & $\!\!$10.26~\  0.11  \\ 
\hline
\label{resmean}
\end{tabular}
\end{table}

\subsection{Burst strengths and ages}
\label{strength}

\begin{figure}
\psfig{bbllx=192.,bblly=395.,bburx=380.,bbury=787.,file=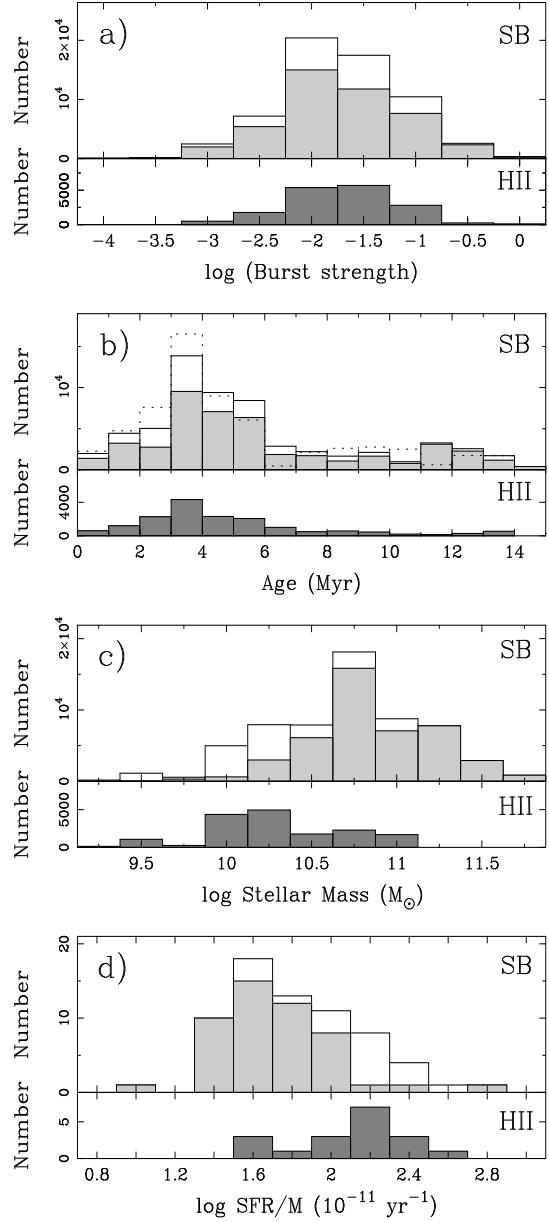,width=8cm,angle=0}
\caption{Frequency histograms of the derived physical properties; {\bf
a)} Burst strength, {\bf b)} age, {\bf c)} $K$-band stellar mass, and
{\bf d)} {\it specific} SFR (SFR per unit mass). {\it Upper panels}
show the histograms for the whole sample ({\it open histograms}) and
for the SB-{\it like} objects ({\it light-grey filled histograms}).
{\it Lower-panels} are the frequency histograms of the HII-{\it like}
galaxies ({\it dark-grey filled histograms}).}
\label{finhisto}
\end{figure}

In Table~\ref{tablafin} we give mean burst strengths and ages for the
individual solutions with probability higher than 20~per cent. Errors
given are the standard deviations of the data points in each solution.
For the stellar masses, the error related with the uncertainty in the
$K$-band absolute magnitude determination is also given (in
parenthesis). Using these probability distributions we have derived the
burst strength, age, mass and metallicity frequency histograms for the
whole sample as well as for the SB-{\it like} and HII-{\it like}
galaxies (see Figures~\ref{finhisto}a--d). The number of points in the
$y$ axis of these figures corresponds to the number of Monte Carlo test
particles with a given burst strength, age, mass or metallicity within
the accepted high-probability solutions.

This analysis yields a typical burst strength of 2$\times$10$^{-2}$
with approximately 90~per cent of the sample having burst strengths
between 10$^{-3}$ and 10$^{-1}$. Only seven objects in the sample show
burst strengths higher than 10$^{-1}$ with a probability larger than
50~per cent, UCM0003$+$2200, UCM0145$+$2519, UCM1257$+$2808,
UCM1259$+$3011, UCM1308$+$2958, UCM1432$+$2645 and UCM1440$+$2511.

Although the properties of the local star-forming galaxies seem to be
well reproduced with an episodic star formation history (see also
AH96), some of these objects may have evolved under more constant star
formation rates (Glazebrook et al. 1999; Coziol 1996). In those objects
the instantaneous burst assumption could yield very high burst
strengths. 

The burst strength histograms shown in Figure~\ref{finhisto}a give
typically larger burst strengths for the HII-{\it like} objects,
especially for the DHIIH and BCD type galaxies, than for the SB-{\it
like} (see also Table~\ref{resmean}). This segregation in burst
strength is probably related to the difference in mean EW(H$\alpha$)
pointed out in section~\ref{mean} (see Table~\ref{datamean}). In
Table~\ref{resmean} we also show the burst strengths and ages derived
under the unrealistic assumption that the continuum extinction is as
high as that measured for the ionized gas (see
section~\ref{minimization}).

The distribution of the burst ages is shown in Figure~\ref{finhisto}b.
Since the probability of detection increases with EW(H$\alpha$) in
objective-prism surveys (see Garc\'{\i}a-Dab\'o et al. 1999), and the
EW(H$\alpha$) continuously decreases with the the burst age, the
number of objects detected with old burst ages is expected to be lower
than with young ages, as observed.  This behaviour is observed at ages
older than 4\,Myr, both for the SB and HII-{\it like} galaxies.

However, one would expect a reasonably flat distribution in the number
of objects with young ages if the sample selection depended only on the
H$\alpha$ equivalent width. But other factors such as the H$\alpha$
flux and continuum luminosity play an important role (see
Garc\'{\i}a-Dab\'o et al. 1999). Moreover, in our models we have
estimated the H$\alpha$ luminosity ($L_{\mathrm{H}\alpha}$) from the
number of Lyman continuum photons (Brocklehurst 1971) assuming that no
ionizing photons escape from the galaxies.  If some Lyman photons
escape, the predicted H$\alpha$ luminosity would be lower and the
derived ages could be significantly younger.  Recent studies estimate
the fraction of Lyman photons escaping from starburst galaxies to be
about 3~per cent \cite{leitherer95}.  Bland-Hawthorn \& Maloney
\shortcite{bland97} estimated this quantity to be about 5~per cent for
the Milky Way.  Another feasible explanation could be that a
significant fraction of these Lyman photons is absorbed by dust within
the ionized gas (see, e.g., Armand et al. 1996). Both mechanisms would
produce lower H$\alpha$ equivalent widths than those predicted by the
standard {\it super-ionizing} models, and could explain the paucity of
young star-forming bursts in Figure~\ref{finhisto}b. In this figure
({\it dotted line}) we also show the age distribution obtained
assuming that 25~per cent of the Lyman continuum photons are
missing. This distribution yields a larger number of objects at ages
younger than 3\,Myr, and a very steep decay at ages older than
4--5\,Myr.

Finally, Bernasconi \& Maeder \shortcite{bernasconi} have
argued that, during the first 2--3\,Myr in the main-sequence, stars more
massive than 40\,M$_{\sun}$, are still accreting mass embedded in the
molecular cloud, and do not contribute to the ionizing radiation.
Therefore, due to this reduction in the number of Lyman photons, the
predicted H$\alpha$ equivalent widths below 2-3\,Myr will be
significantly lower and the ages deduced for the bursts should be
younger.

\subsection{Total stellar masses}
\label{mass}

In order to determine the total galaxy stellar mass we have assumed
that the burst strengths and mass-to-light ratios derived from our
models at three disk scale-length apertures are representative of the
galaxy global properties.  Thus, using these $K$-band mass-to-light
ratios and the total $K$-band absolute magnitudes we have obtained
stellar masses for the whole sample. 

\begin{figure}
\psfig{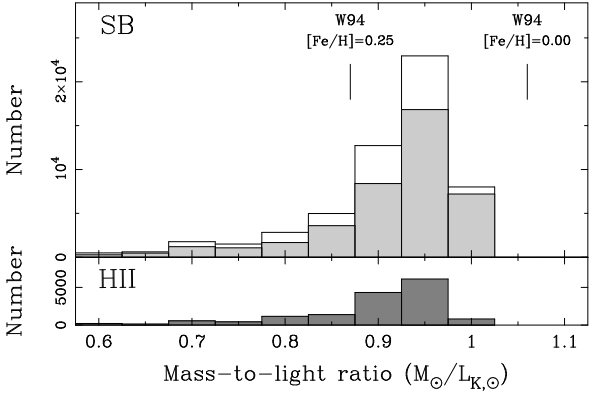}
\caption{$K$-band mass-to-light ratio frequency histogram. The {\it upper panel}
shows the distribution for the whole sample ({\it open histogram}) and
SB-{\it like} objects ({\it light-grey filled histogram}). The {\it
lower-panel} shows the frequency histogram of the HII-{\it like}
galaxies ({\it dark-grey filled histogram}). Mass-to-light ratios for
model ellipticals with different metallicities are taken from worthey
(1994, w94).}
\label{mlratio}
\end{figure}

The inferred galaxy stellar masses derived depend, in principle, on
four quantities: the galaxy $K$-band absolute magnitude, burst
strength, and the mass-to-light ratios of the burst and the old
underlying population. Since the derived burst strengths are very low
($\sim$10$^{-2}$), the total mass-to-light ratios are dominated by the
old stellar component. In fact, the ratio of the $K$-band luminosity
of the old and young stellar populations is $\sim$20 for $t$=4\,Myr, 4
for $t$=8\,Myr and 7 for $t$=15\,Myr (for $Z$=$Z_{\sun}$).  Moreover,
the absolute age of the old stellar component has a very small effect
on the $K$-band mass-to-light ratio: there is only a 0.1\,dex
difference between 10\,Gyr and 15\,Gyr for solar metallicity, and the
difference is even lower in the case of sub-solar metallicity models.
In Figure~\ref{mlratio} we show that the derived $K$-band stellar
mass-to-light ratios span a very narrow range.  Although statistically
the SB- and HII-{\it like} mass-to-light ratio distributions are
different with a probability of 95.3~per cent (from a K-S test), the
difference in absolute value is only minor: the median mass-to-light
ratios are 0.93\,M$_{\sun}$/L$_{K,\sun}$ for the whole sample, and
0.93\,M$_{\sun}$/L$_{K,\sun}$ and 0.91\,M$_{\sun}$/L$_{K,\sun}$, for
the SB and HII-{\it like} objects respectively. Consequently, the
derived mass values mainly depend on the $K$-band absolute magnitude.
In the {\it upper-panel\/} of Figure~\ref{mlratio} we also show the
range in $K$-band mass-to-light ratios given by Worthey
\shortcite{worthey} for 12\,Gyr old modeled ellipticals.  Thus, we can
conclude that the $K$-band luminosity is a very good tracer of the
stellar mass for both old stellar populations and local star-forming
objects.

The distribution of stellar masses is shown in Figure~\ref{finhisto}c.
This frequency histogram indicates that a typical star forming galaxy
in our Local Universe has a stellar mass of about
5$\times$10$^{10}$\,M$_{\sun}$.  This value is somewhat lower than the
stellar mass expected for a local L$^{*}$ galaxy. Assuming
M$^{*}_{K}$=$-$25.1 (for H$_0$=50\,km\,s$^{-1}$\,Mpc$^{-1}$; Mobasher,
Sharples and Ellis, 1993) and a $K$-band mass-to-light ratio of
1\,M$_{\sun}$/L$_{K,\sun}$ \cite{heraudeau}, the stellar mass inferred
for an L$^{*}$ galaxy is about 2$\times$10$^{11}$\,M$_{\sun}$. Thus,
star-forming galaxies in the local universe are typically a factor~4
less massive than L$^{*}$ galaxies.

In addition, a clear offset between the stellar mass histograms of the
SB and HII-{\it like} objects is seen in Figure~\ref{finhisto}c. The
distributions of their stellar masses are centred at
7$\times$10$^{10}$ and 2$\times$10$^{10}$\,M$_{\sun}$ respectively
(Figure~\ref{finhisto}c). This difference is even more significant,
about 1\,dex, when DHIIH and BCD spectroscopic types ({\it Dwarfs} in
Table~\ref{resmean}) and SBN galaxies are compared.  A K-S test
analysis of the SB-{\it like} and HII-{\it like} objects indicates
that these two samples come from different age, burst strength and
stellar mass distributions with probabilities 98.8, 77.1 and 99.9 per
cent respectively.

In Table~\ref{resmean} we also present the mean properties that would be
obtained using $E(B-V)_{\mathrm{continuum}}=E(B-V)_{\mathrm{gas}}$. In
this case, although we obtain important differences in age and burst
strength, very similar stellar masses are derived since the stellar
mass depends mainly on the $K$-band magnitude, only weakly affected by
extinction.

\begin{table}
\caption[]{Mean and median log($L_{\mathrm{H}\alpha}$/SFR) ratios obtained for different metallicities.}
\begin{tabular}{rccc}
Metallicity  & \multicolumn{3}{c}{log ($L_{\mathrm{H}\alpha}$/SFR)} \\
\hline
 $Z$ & Mean & Median & Std.dev. \\
\hline 
1/50 Z$_{\sun}$  & 40.19 & 40.09 & 0.58\\
1/5 Z$_{\sun}$   & 40.38 & 40.35 & 0.36\\
2/5 Z$_{\sun}$   & 40.31 & 40.27 & 0.29\\
Z$_{\sun}$       & 40.24 & 40.29 & 0.31\\
2 Z$_{\sun}$     & 40.19 & 40.21 & 0.34\\
\hline
All $Z$            & 40.23 & 40.23 & 0.44\\
\hline
\label{fig6t}
\end{tabular}
\end{table}

\begin{table}
\caption[]{Mean SFR and {\it specific} SFR for the sample, together with 
the corresponding standard deviations of the mean. {\it Dwarfs\/}
includes DHIIH and BCD spectroscopic type galaxies.}
\begin{tabular}{lrcc}
 & n & log(SFR) \ \ $\sigma$ & log(SFR/M) \ \ $\sigma$ \\
\hline 
 &   & (M$_{\sun}$\,yr$^{-1}$) &    (10$^{-11}$\,yr$^{-1}$)     \\
\hline
Total & 67 &       \ \ \ \ 1.52\ \ \ \   \ \ 0.05 & \ \ \ \ 1.81\ \ \ \ \ \ \ \ 0.04 \\ 
SBN   & 41 &       \ \ \ \ 1.64\ \ \ \   \ \ 0.06 & \ \ \ \ 1.73\ \ \ \ \ \ \ \ 0.05 \\
DANS  &  8 &       \ \ \ \ 1.21\ \ \ \   \ \ 0.09 & \ \ \ \ 1.66\ \ \ \ \ \ \ \ 0.09 \\ 
HIIH  & 13 &       \ \ \ \ 1.60\ \ \ \   \ \ 0.08 & \ \ \ \ 2.16\ \ \ \ \ \ \ \ 0.07 \\
{\it Dwarfs} & 5 & \ \ \ \ 0.82\ \ \ \   \ \ 0.09 & \ \ \ \ 1.88\ \ \ \ \ \ \ \ 0.14 \\
\hline
SB    & 49 &       \ \ \ \ 1.57\ \ \ \   \ \ 0.05 & \ \ \ \ 1.72\ \ \ \ \ \ \ \ 0.04 \\  
HII   & 18 &       \ \ \ \ 1.38\ \ \ \   \ \ 0.10 & \ \ \ \ 2.08\ \ \ \ \ \ \ \ 0.07 \\  
\hline
\label{tablasfrs}
\end{tabular}
\end{table}

\begin{figure}
\psfig{bbllx=165.,bblly=235.,bburx=330.,bbury=744.,file=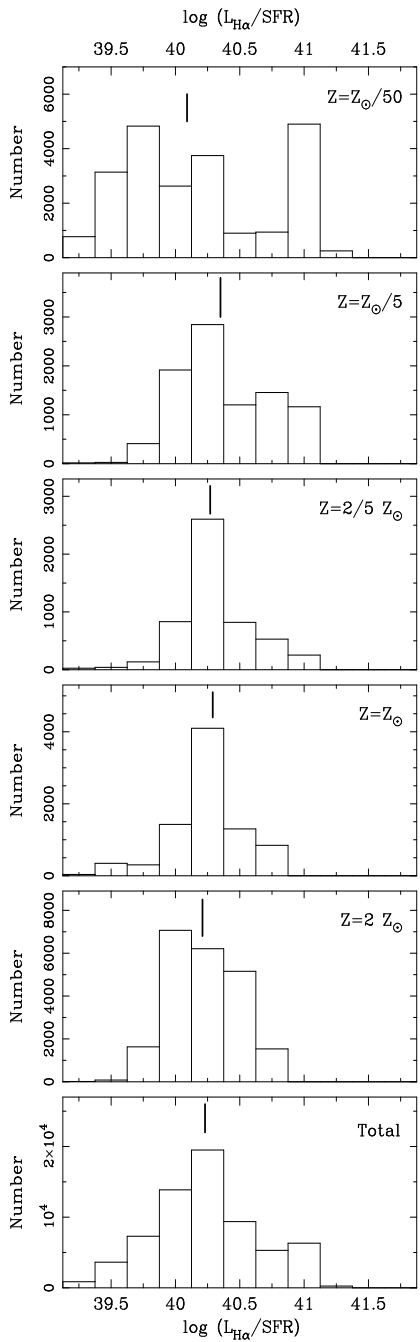,height=18.cm,angle=0}
\caption{log$\,$($L_{\mathrm{H}\alpha}$/SFR) frequency histograms for different metallicities and for the whole distribution. Thick marks give the position of the distribution median values.}
\label{fig6}
\end{figure}

\begin{figure}
\psfig{bbllx=32.,bblly=232.,bburx=300.,bbury=750.,file=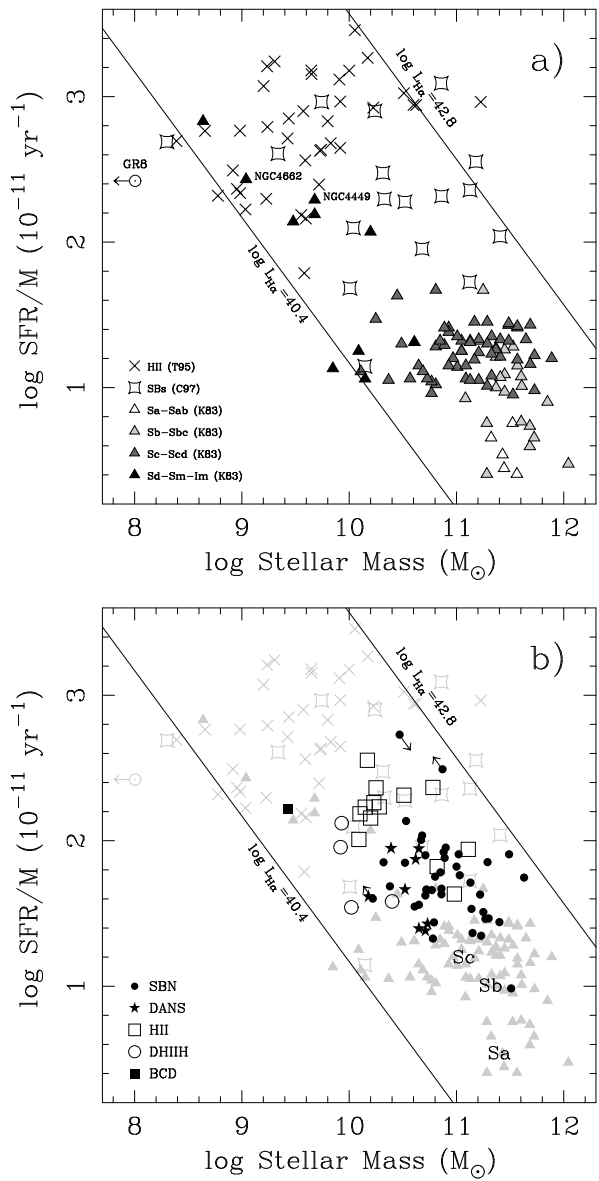,height=16.cm,width=8.2cm,angle=0}
\caption{Stellar mass (in M$_{\sun}$) vs. {\it specific} SFR (in 10$^{-11}$\,yr$^{-1}$). 
{\bf a)} Dynamical masses from Telles (1995), converted to stellar
masses using a correction factor of 0.6\,dex. Starburst galaxies from
Calzetti (1997b) are also plotted. Data for spiral galaxies have been
taken from Kennicutt (1983). {\bf b)} SBN, DANS, HII, DHIIH and BCD
UCM galaxies as classified by Gallego et al. (1996) are
plotted. Straight lines define the limits in the H$\alpha$ luminosity
function of the UCM survey.}
\label{sfrs}
\end{figure}

\subsection{Star formation rates}
\label{sfr}

Since the star formation activity in the UCM galaxies is better described
as a succession of episodic star formation events rather than
continuous star formation (see also AH96), the current star formation
rate (SFR) is not a meaningful quantity: the latest star formation
event might have finished in many of the galaxies, and their current
SFR would be zero. However, these galaxies have substantial H$\alpha$
luminosities, and it is accepted that the H$\alpha$ luminosity is a
good measurement of the current SFR. In AH96 we showed that this is
true, in a statistical sense, for a population of galaxies undergoing a
series of star formation events, and we defined an `effective'
present-day SFR which coincides with the SFR we would derive if 
the galaxies were forming stars at a constant rate, producing the
same mass in new stars as the ensemble of all the star-formation
episodes (see AH96 for details). Here we will follow the same approach. 

When estimating star formation rates (SFRs) in AH96 we used the BC93
models and a Salpeter IMF. In the present work, we have used the
updated BC96 models with a Scalo IMF. Since the number of Lyman photons
(N$_{\mathrm{Ly}}$) predicted by the old models is about 0.94\,dex
higher than that predicted by the new ones (for solar metallicity and
ages lower than 16\,Myr), we need to re-compute the relation between
the H$\alpha$ luminosity, $L_{\mathrm{H}\alpha}$, and star formation
rate.  In addition, we will investigate the change produced in this
relation using different metallicity models.

In order to compute the $L_{\mathrm{H}\alpha}$/SFR ratio, we have used
a very similar procedure to that employed by AH96: we simulated a
population of galaxies undergoing random bursts of star-formation and
computed their total H$\alpha$ luminosities and the mass in
newly-formed stars. However, instead of considering a uniform age and
burst strength probability distribution, we have considered the burst
strength, age and metallicity distributions for our galaxy sample.  We
used 67$\times$1000 points in order to reproduce this distribution in
our Monte Carlo simulations.  The SFR was computed as the ratio
between the stellar mass produced in the burst, i.e. $b$$\times$M, and
the maximum age for which we could have detected the galaxy in the UCM
sample, that is, the time while EW(H$\alpha$)$>$20\AA\
\cite{gallego95a}. The $L_{\mathrm{H}\alpha}$/SFR ratios obtained are
shown in Figure~\ref{fig6} for different metallicities and for the
total ($t$,$b$,$Z$) distribution. The mean, median, and standard
deviation values are given in Table~\ref{fig6t}.

Since the changes in this ratio for different metallicity models are
quite small, we have adopted the median value of the whole distribution
in order to determine the 'effective' SFR of the galaxies from
our sample. The difference between the value adopted here and that of
AH96 is about 1\,dex, which is very close to the difference in the
number of Lyman photons predicted by the BC93 and BC96 models, as
expected.
 
Therefore, we have evaluated the current 'effective' SFR using the
expression
\begin{equation}
\mathrm{SFR} = {L_{\mathrm{H}\alpha} \over 1.7\times10^{40} \mathrm{erg}\,\mathrm{s}^{-1}} \ \ \ \ \mathrm{M}_{\sun}\,\mathrm{yr}^{-1}
\label{eq_sfr}
\end{equation}
This expression assumes that every Lyman photon emitted effectively
ionizes the surrounding gas.  If, however, as is suggested in
Section~\ref{strength}, we consider a fraction of non-ionizing Lyman
photons of 25~per cent, the star formation rates computed should be
0.1\,dex higher.

{\it Specific} star formation rates (SFR per unit mass; Guzm\'an et
al. 1997) have been obtained using these SFR values and the stellar
masses given by the highest probability solution cluster in
Table~\ref{tablafin}. The mean SFR and {\it specific} SFR for SB-{\it
like}, HII-{\it like}, and whole sample are given in
Table~\ref{tablasfrs} (see also Figure~\ref{finhisto}d).

The {\it specific} SFR vs. stellar mass diagram is shown in
Figure~\ref{sfrs} (see Guzm\'an et al. 1997). In panel-{\bf a} we show
the stellar masses and star formation rates per unit mass for three
reference samples. We have included the sample of Kennicutt (1983, K83
hereafter), taking the H$\alpha$ and $B$-band luminosities given by
K83 and the stellar mass-to-light ratios of Faber \& Gallagher
\shortcite{faber}. In addition, the sample of HII-galaxies
of Telles \shortcite{telles95} was included, after converting virial
masses to stellar masses using a correction of 0.6\,dex (Gallego et
al. 1999, in preparation) and assuming the H$\beta$-to-H$\alpha$
luminosity ratios used by Guzm\'an et al. \shortcite{guzman97} for
this sample. Masses and {\it specific} SFRs for the Calzetti
\shortcite{calzetti97b} sample are also shown. In this case, stellar
masses were inferred subtracting the HI mass from the dynamical mass
measured. The SFR values for the Calzetti \shortcite{calzetti97b}
sample were obtained from their Br$\gamma$ luminosities assuming
$L_{\mathrm{H}\alpha}$=102.8$\times$$L_{\mathrm{Br}\gamma}$
(Osterbrook 1989, for $T_{e}$=10$^{4}$\,K and
$n_{e}$=100\,cm$^{-3}$). Finally, the dwarf irregular galaxy GR8
\cite{reaves} was included. Its H$\alpha$ luminosity was
obtained from the H$\beta$ luminosity given by Gallagher, Hunter \&
Bushouse \shortcite{gallagher89}, assuming
$L_{\mathrm{H}\alpha}$/$L_{\mathrm{H}\beta}$=2.86, and its stellar
mass, 3.2$\times$10$^{6}$\,M$_{\sun}$, from Carignan, Beaulieu \&
Freeman \shortcite{carignan}. The limits in the H$\alpha$ luminosity
function of Gallego et al. \shortcite{gallego95b},
10$^{40.4}$--10$^{42.8}$\,erg\,s$^{-1}$, are also drawn.

Figure~\ref{sfrs} shows that the UCM sample clearly represents a bridge
between relaxed spiral galaxies and the most extreme HII galaxies from
Telles \shortcite{telles95}, that is, Sp$\rightarrow$SB-{\it
like}$\rightarrow$HII-{\it like}$\rightarrow$HII galaxies. In fact,
some of the HII galaxies from Telles \shortcite{telles95} have very
similar properties to those of the less massive HII-{\it like} UCM
galaxies, mainly DHIIH and BCD spectroscopic types, very rare in our
sample (see section~\ref{completeness}). In addition, most of the SBN
type UCM galaxies seem to be normal late-type spirals with enhanced
star formation. This star formation enhancement is about a factor of
three, and is due to the ongoing nuclear starburst.  Thus, the range in
{\it specific} SFR spanned by the population of the star-forming
galaxies that dominate the SFR in the local universe is
(10--10$^{3}$)$\times$10$^{-11}$\,yr$^{-1}$, from the local relaxed
spirals to the most extreme HII galaxies. In fact, this range is not
very different from that obtained by Guzm\'an et al.
\shortcite{guzman97} for a sample of intermediate/high-$z$ compact
galaxies from the HDF. The high {\it specific} SFR region, where the
HII galaxies from Telles \shortcite{telles95} are placed, is not very
well covered by our sample due to the scarcity of very low-luminosity
objects, basically DHIIH and BCD spectroscopic type galaxies, relative
to the UCM whole sample.

\section{Summary}
\label{summary}

Using new nIR observations and published optical data for 67 galaxies
from the Universidad Complutense de Madrid (UCM) survey, we have
derived the main properties of their star-forming events and underlying
stellar populations. This sample represents about 35~per cent of the
UCM galaxies covering the whole range of absolute magnitudes, H$\alpha$
luminosities and equivalent widths spanned by the survey. Burst
strengths and ages, stellar masses, stellar mass-to-light ratios
and, to a certain extent, metallicities, have been obtained by
comparing the observed $r-J$ and $J-K$ colours, $K$-band magnitudes, and
H$\alpha$ equivalent widths and luminosities with those predicted by
evolutionary synthesis models.  The comparison of the observations with
the model predictions was carried out using a maximum-likelihood
estimator in combination with Monte Carlo simulations which take into
account the observational uncertainties. Our main results are: 

\begin{enumerate}

       \item The star-forming galaxies in the UCM sample (used to
        determine the SFR density of the local universe), show typical
        burst strengths of about 2~per cent and stellar masses of
        5$\times$10$^{10}$\,M$_{\sun}$. The current star
        formation in these galaxies is taking place in discrete 
        star formation events rather than in a continuous fashion. 
	If this is typical of the past star-formation history in
	the galaxies, many of such star formation events would be 
	necessary to build up their stellar mass. However, our observations
	provide very little information on star-formation episodes 
	that took place before the current one.

        \item We have identified two separate classes of star-forming
        galaxies in the UCM sample: SB-{\it like} and HII-{\it like}
        galaxies. Within the HII-{\it like} class the DHIIH and BCD
        spectroscopic type galaxies, i.e. {\it dwarfs}, constitute the
        most extreme case. The mean burst strength deduced for the
        SB-{\it like} galaxies is about a 25~per cent lower than for
        the {\it dwarf} HII-{\it like} galaxies. The average stellar
        mass is an order of magnitude larger in the former than in the
        latter. The SB-{\it like} galaxies are relatively massive
        galaxies where the current star formation episode is a minor
        event in the build up of their stellar masses, while HII-{\it
        like} galaxies are less massive systems in which the present
        star formation could dominate in some cases their observed
        properties and contributes to a greater extent to their
        stellar population.

        \item Because of the low burst strengths inferred, $K$-band
        luminosity is dominated by the old stellar populations, and
        the $K$-band stellar mass-to-light ratio is almost the same
        (within $\sim$20~per cent) for all the galaxies. Thus, the
        $K$-band luminosity is a very good estimator of the stellar
        mass for typical star-forming galaxies.

        \item The average SFR of the galaxies is log(SFR)$\simeq$1.5,
        with the SFR expressed in M$_{\sun}$\,yr$^{-1}$, and it is
        similar for the SB-{\it like} and the HII-{\it like}
        galaxies. However, since the latter are typically less
        massive, their specific SFR (SFR per unit stellar mass) is
        significantly larger, in a factor 2.3, than that of the
        former.

        \item The UCM galaxies represent a bridge in {\it specific}
        SFR between relaxed spirals and extreme HII galaxies. The
        range in {\it specific} star formation rate spanned by the
        local star-forming galaxies,
        (10--10$^{3}$)$\times$10$^{-11}$\,yr$^{-1}$, is very similar
        to that observed in higher redshift objects.

  \end{enumerate}

\section*{APPENDIX: Analysis of the space of solutions}

In this appendix we briefly describe the analysis performed onto the
($t$,log\,$b$,log\,$Z$) space of solutions. For each galaxy we have
10$^{3}$ ($t$,log\,$b$,log\,$Z$) points that correspond to the
10$^{3}$ points generated in the
($r-J$,$J-K$,2.5$\times$log\,EW(H$\alpha$)) space using a Monte Carlo
simulation method.

The mean values ($<t>$,$<$log\,$b$$>$,$<$log\,$Z$$>$) are primary
indicators of the best ($t$,log\,$b$,log\,$Z$) solution, and their
standard deviations
($\sigma_t$,$\sigma_{\mathrm{log}\,b}$,$\sigma_{\mathrm{log}\,Z}$)
could be taken as estimators of the deviation of the data. However,
due to the well known age-metallicity and age-burst strength
degeneracies these standard deviations are not representative of the
distribution of these solutions in the ($t$,log\,$b$,log\,$Z$)
space. Fortunately, these degenaracies do not span the whole range in
age, burst strength and metallicity given by the models, being
relatively well constrained by the
($r-J$,$J-K$,2.5$\times$log\,EW(H$\alpha$)) data.

\begin{figure}
\psfig{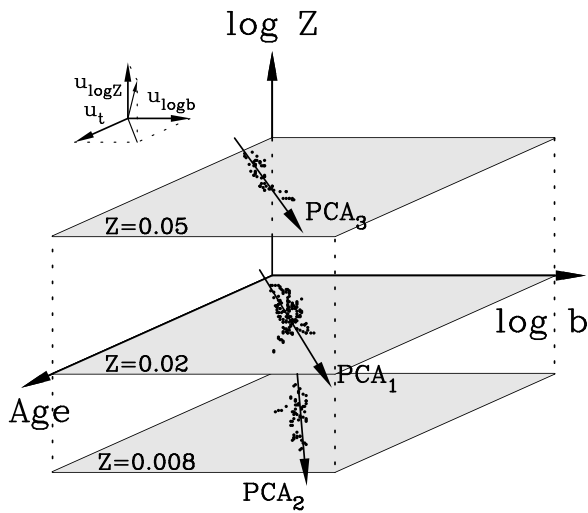}
\caption{Hypothetical distribution of ($t$,log\,$b$,log\,$Z$) solutions. PCA1, PCA2, PCA3 are the principal components for each of the three solution clusters.}
\label{fig8}
\end{figure}

Therefore, we have studied the clustering of the
($t$,log\,$b$,log\,$Z$) solutions for each individual galaxy. We have
used for this analysis a single linkage hierarchical clustering method
\cite{murtagh}. First, {\bf (1)} we determine the distances between
every couple of solutions, which represents a total of $N\times(N-1)/2$
dissimilarities (=distances), being N the number of solutions. The
dissimilarity between the elements $j$ and $k$, $d_{j,k}$, is defined
as
\begin{equation}
d_{j,k}^{2} = \sum_{i=1}^{n} (x_{ij} - x_{ik})^{2}
\end{equation}
The matrix of dissimilarities is known as {\it dendogram}. Then, {\bf
(2)} we find the smallest dissimilarity, $d_{ik}$. These points,
$i$ and $k$, {\bf (3)} are therefore agglomerated and replaced with a
new point, $i \cup k$, and the dissimilarities updated such that, for
all objects $j \neq i,k$,
\begin{equation}
d_{i\cup k,j} = min\ \{d_{i,j},d_{k,j}\} 
\end{equation}
Then, {\bf (4)} the dissimilarities $d_{i,j}$ and $d_{k,j}$, for all $j$,
are deleted, as these are no longer used. Finally, we return to step
{\bf (1)} after reducing the dimension of the dissimilarities matrix
and the number of clusters. Finally, we recover the last three
clusters of solutions.

The clustering pattern obtained is basically produced by the
discretization in metallicity of the original BC96 evolutionary
synthesis models. Now, we analyze the solutions within each solution
cluster. In this case, the discretizations in burst strength and age
are comparable and a Principal Component Analysis (PCA hereafter) is
the most suitable choice (0.04\,dex in burst strength and
$\sim$0.05\,dex in age).
 
The PCA basically determines, in a R$^{n}$ data array, the set of $n$
orthogonal axes that better reproduces our data distribution. The
first new axis, i.e. the principal component, will try to go as close
as possible through all the data points, describing the larger
fraction of the data variance. Figure~\ref{fig8} shows the principal
component for each of the three solution clusters of a hypothetical
($t$,log\,$b$,log\,$Z$) distribution.

Formally, following the PCA (see, e.g., Morrison 1976), {\bf (1)} we
construct the variance-covariance and the correlation matrix of the
sample, being the $(j,k)^{th}$ term of these matrixes, respectively, 
\begin{eqnarray}
c_{jk} = {1 \over n} \sum_{i=1}^{n} (r_{ij} - \overline{r}_j) (r_{ik}-\overline{r}_k)\\
\rho_{jk} = {1 \over n} \sum_{i=1}^{n} {{(r_{ij} - \overline{r}_j) (r_{ik}-\overline{r}_k)} \over {s_{j} s_{k}}}
\end{eqnarray}
where
\begin{equation}
s_{j}^{2} = {1 \over n} \sum_{i=1}^{n} (r_{ij} - \overline{r}_j)^{2}.
\end{equation}
Then, {\bf (2)} solving the eigenvector equation, $\rho u$=$\lambda
u$, we obtain the eigenvalues and eigenvectors of the correlation
matrix. The ratio between an eigenvalue and the sum of all the
eigenvalues, $\lambda_i/\sum_{i=1}^{n}\lambda_i$, gives us the
contribution of the new axis, determined by the corresponding
eigenvector, to the total data variance. Therefore, the eigenvector
with higher eigenvalue is the principal component and will indicate
which is the dominant degeneracy inside each solution cluster.

\section*{ACKNOWLEDGMENTS}
This work is based on observations obtained at the Lick Observatory,
operated by the University of California and on observations collected
at the German-Spanish Astronomical Centre, Calar Alto, Spain, operated
by the Max-Planck Institute fur Astronomie (MPIE), Heidelberg, jointly
with the Spanish Commission for Astronomy. It is also partly based on
observations made with the Isaac Newton Telescope operated on the
island of La Palma by the Royal Greenwich Observatory in the Spanish
Observatorio del Roque de los Muchachos of the Instituto de
Astrof\'{\i}sica de Canarias.

A. Gil de Paz thanks the Institute of Astronomy of the University of Cambridge
for all the facilities and support during his stay there. J.  Gallego and A.
Gil de Paz acknowledge the invitation, hospitality and facilities provided
during the 3rd Guillermo Haro Workshop, included in the Guillermo Haro
Programme at the INAOE (Mexico). We also thank C.E.  Garc\'{\i}a-Dab\'o, C.
S\'anchez Contreras and R. Guzm\'an for stimulating conversations and the
referee Dr. M. Edmunds for his useful comments and suggestions. A. Gil de Paz
acknowledges the receipt of a {\it Formaci\'on del Profesorado Universitario}
fellowship from the Spanish Ministry of Education. A. Arag\'on-Salamanca
acknowledges generous financial support from the Royal Society. A.
Alonso-Herrero was supported by NASA on grant NAG 5-3042. This research was
also supported by the Spanish Programa Sectorial de Promoci\'on General del
Conocimiento under grants PB96-0610 and PB96-0645.  This work was partially
carried out under the auspices of EARA, a European Association for Research in
Astronomy, and the TMR Network on Galaxy Formation and Evolution funded by the
European Commission.

\bsp

\label{lastpage}

\end{document}